\documentclass[acmtog, nonacm]{acmart}

\AtBeginDocument{%
  }

\newcommand{\beginsupplement}{%
        \setcounter{table}{0}
        \renewcommand{\thetable}{S\arabic{table}}%
        \setcounter{figure}{0}
        \renewcommand{\thefigure}{S\arabic{figure}}%
        \setcounter{section}{0}
        \renewcommand{\thesection}{S\arabic{section}}%
     }

\let\maketitlesup\maketitle
\usepackage{xpatch}
\xpatchcmd{\maketitlesup}{\@mkteasers}{}{}{}
\xpatchcmd{\maketitlesup}{\@mkabstract}{}{}

\usepackage[ruled]{algorithm2e} 

\SetAlFnt{\small}
\SetAlCapFnt{\small}
\SetAlCapNameFnt{\small}
\SetAlCapHSkip{0pt}
     
\copyrightyear{2024}
\acmYear{2024}

\usepackage{booktabs} 
\usepackage{placeins} 

\usepackage{soul}
\usepackage{xcolor}

\usepackage{xcolor}

\usepackage{graphicx}
\usepackage{subcaption}
\usepackage{float}
\usepackage{siunitx}

\begin{document}

\title{HoloChrome: Polychromatic Illumination for Speckle Reduction in Holographic Near-Eye Displays}


\author{Florian Schiffers}
\affiliation{%
  \institution{Reality Labs Research, Meta}
  \city{Redmond}
  \state{WA}
  \country{USA}}
\affiliation{%
  \institution{Northwestern University}
  \city{Evanston}
  \state{IL}
  \country{USA}}
\email{florianschiffers@gmail.com}
\author{Grace Kuo}
\affiliation{%
  \institution{Reality Labs Research, Meta}
  \city{Redmond}
  \state{WA}
  \country{USA}}
\author{Nathan Matsuda}
\affiliation{%
  \institution{Reality Labs Research, Meta}
  \city{Redmond}
  \state{WA}
  \country{USA}}
\author{Douglas Lanman}
\affiliation{%
  \institution{Reality Labs Research, Meta}
  \city{Redmond}
  \state{WA}
  \country{USA}}
\author{Oliver Cossairt}
\affiliation{%
  \institution{Reality Labs Research, Meta}
  \city{Redmond}
  \state{WA}
  \country{USA}}
\renewcommand\shortauthors{Schiffers, F. et al}

\authorsaddresses{}


\renewcommand{\shortauthors}{Schiffers, F. et al.}

\begin{abstract}

Holographic displays hold the promise of providing authentic depth cues, resulting in enhanced immersive visual experiences for near-eye applications. However, current holographic displays are hindered by speckle noise, which limits accurate reproduction of color and texture in displayed images. We present HoloChrome, a polychromatic holographic display framework designed to mitigate these limitations. HoloChrome utilizes an ultrafast, wavelength-adjustable laser and a dual-Spatial Light Modulator (SLM) architecture, enabling the multiplexing of a large set of discrete wavelengths across the visible spectrum. By leveraging spatial separation in our dual-SLM setup, we independently manipulate speckle patterns across multiple wavelengths. This novel approach effectively reduces speckle noise through incoherent averaging achieved by wavelength multiplexing. Our method is complementary to existing speckle reduction techniques, offering a new pathway to address this challenge. Furthermore, the use of polychromatic illumination broadens the achievable color gamut compared to traditional three-color primary holographic displays.

Our simulations and tabletop experiments validate that HoloChrome significantly reduces speckle noise and expands the color gamut. These advancements enhance the performance of holographic near-eye displays, moving us closer to practical, immersive next-generation visual experiences.

\end{abstract}

\begin{teaserfigure}  
\vspace{3mm}
  \includegraphics[width=\linewidth]{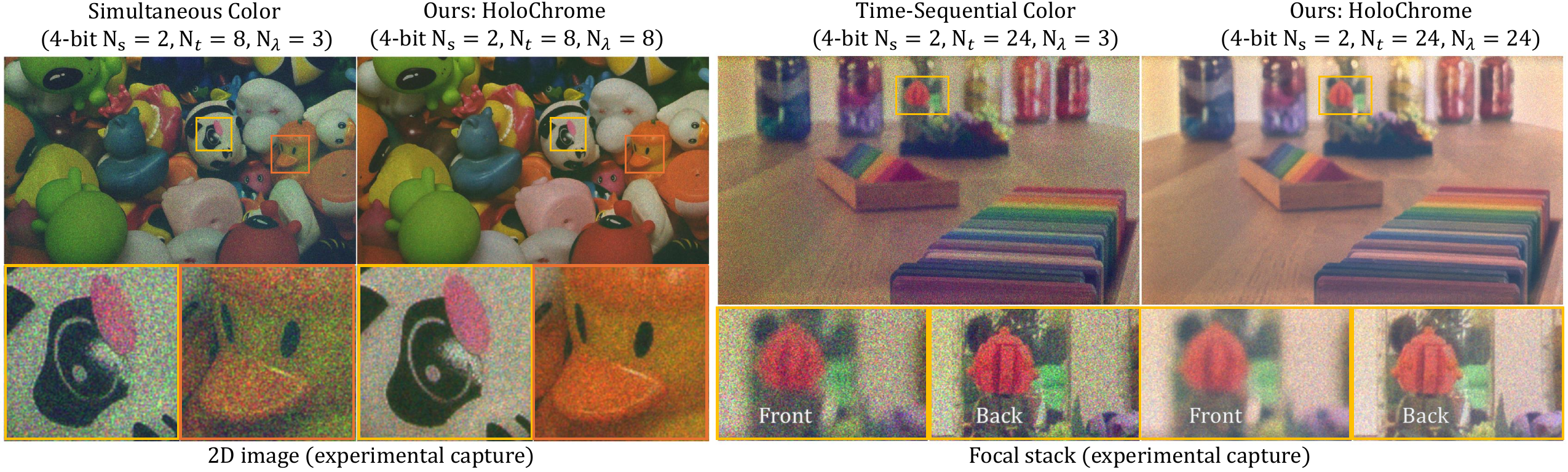}
  \caption{
  \textbf{HoloChrome: a new holographic display architecture specifically designed for speckle reduction}.
  Instead of illuminating the SLM with only three wavelengths, we multiplex polychromatic illumination, which incoherently averages at the detector plane.
  By further using two spatial light modulators with an air gap in between, we break speckle correlations between wavelengths, enabling high-resolution holograms with significantly suppressed speckle noise.
  We experimentally demonstrate significant speckle reduction on both 2D images (left) and 3D focal stacks with natural-looking defocus cues (right). [$N_s$ : number of SLMs, $N_t$ : number of time-multiplexed frames, $N_{\lambda}$ : number of polychromatic wavelengths] 
  }
\vspace{3mm}
\label{fig:teaser_figure}
\end{teaserfigure}


\maketitle

\section{Introduction}

Holographic near-eye displays offer a compelling solution for creating immersive visual experiences with true depth cues and natural image parallax~\cite{maimone2017holographic}.
This field has recently garnered significant interest within the graphics and vision communities, with numerous works published in high-profile journals~\cite{an2020slim,shi2021towards,gopakumar2024full,jang2022waveguide,tseng2024neural}. 
This interest stems from the promise of holographic displays as a comprehensive solution to pass the Visual Turing Test (VTT)~\cite{banks20163d}.
\begin{figure*}[tb]
\centering
\includegraphics[width=1.0\textwidth]{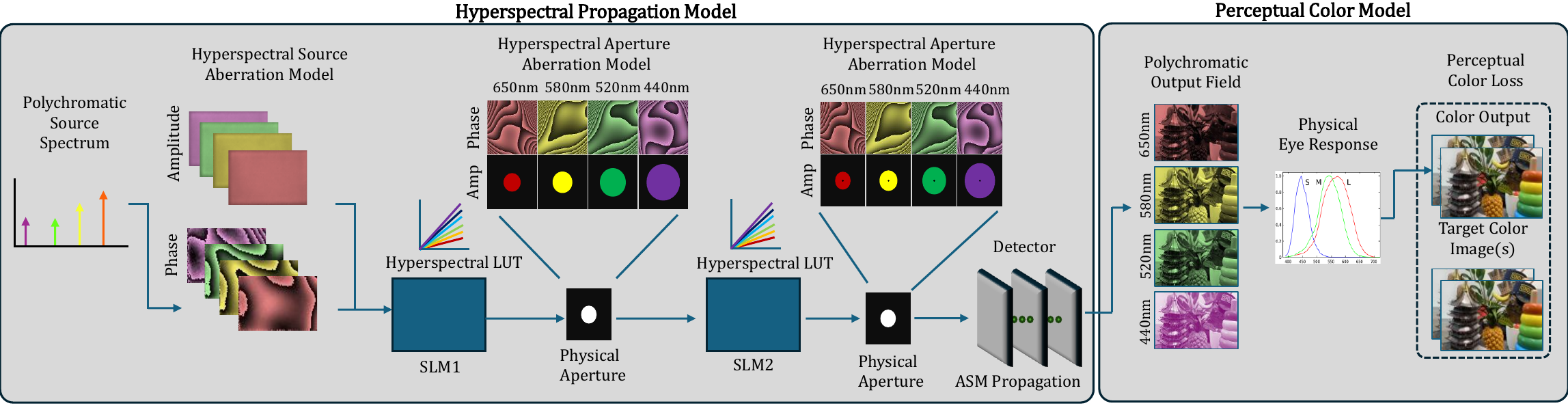}
\Description{Description of the HoloChrome Forward Model concept figure.}
\caption{
\textbf{Hyperspectral propagation model and perceptual optimization framework.}
This figure illustrates the three key components of the proposed HoloChrome framework:
A hyperspectral propagation model (left) is used to generate polychromatic image data cubes that are converted to 3-channel RGB images using perceptual eye response curves, and compared to targets using a perceptual color loss (right).
The hyperspectral propagation model begins with a polychromatic source spectrum that is used to sample a hyperspectral source aberration model with N-discrete wavelengths, generating a polychromatic field with wavelength-dependent amplitude and phase.
These are processed through two spatial light modulators (SLM$_1$ and SLM$_2$) with hyperspectral lookup tables (LUT) that are sampled to create a complex polychromatic aperture to represent frequency domain aberrations.
The polychromatic output field is then measured on a detector after applying wavefront propagation to simulate focal stack capture with a translation stage.
The perceptual response incorporates spectral weighting based on the physical eye response and performs a differentiable color transformation (e.g., XYZ to sRGB) before MSE loss is computed. 
%
}
\label{fig:concept_figure_forward_model}
\vspace{-.1in}
\end{figure*}

An ideal holographic image would consist of the full time-varying optical field incident on a plane over the full duration of the temporal retinal response.
For an incoherent scene, this field is polychromatic with an enormous information density — essentially changing the wavelength, in addition to spatially-varying phase and amplitude every picosecond~\cite{wagner2021intensity}.
When such a field is observed by a human eye, the effect is to average billions of speckle fields every few tens of milliseconds, resulting in a speckle-free perceived image.
The holy grail of holography is to reproduce a close facsimile to this polychromatic hologram with as few degrees of freedom as possible. Unfortunately, the speed of currently available SLMs is many orders of magnitude slower than needed, making such a system far beyond practical implementation.

On the one hand, we would prefer a method to synthesize the exact polychromatic hologram of a scene since it contains the most complete optical representation possible and therefore can be used to pass the VTT. On the other hand, we only care about how our holographic display is perceived by the human eye, which maps this polychromatic information onto just three color primaries.
While we cannot practically synthesize the ideal polychromatic hologram, such a hologram still contains largely redundant information.
Hence, we are left with the problem of how to produce a hologram, within hardware constraints that is also perceptually similar to an ideal polychromatic hologram.   

\begin{figure*}[htb]
\centering
\includegraphics[width=1.0\textwidth]{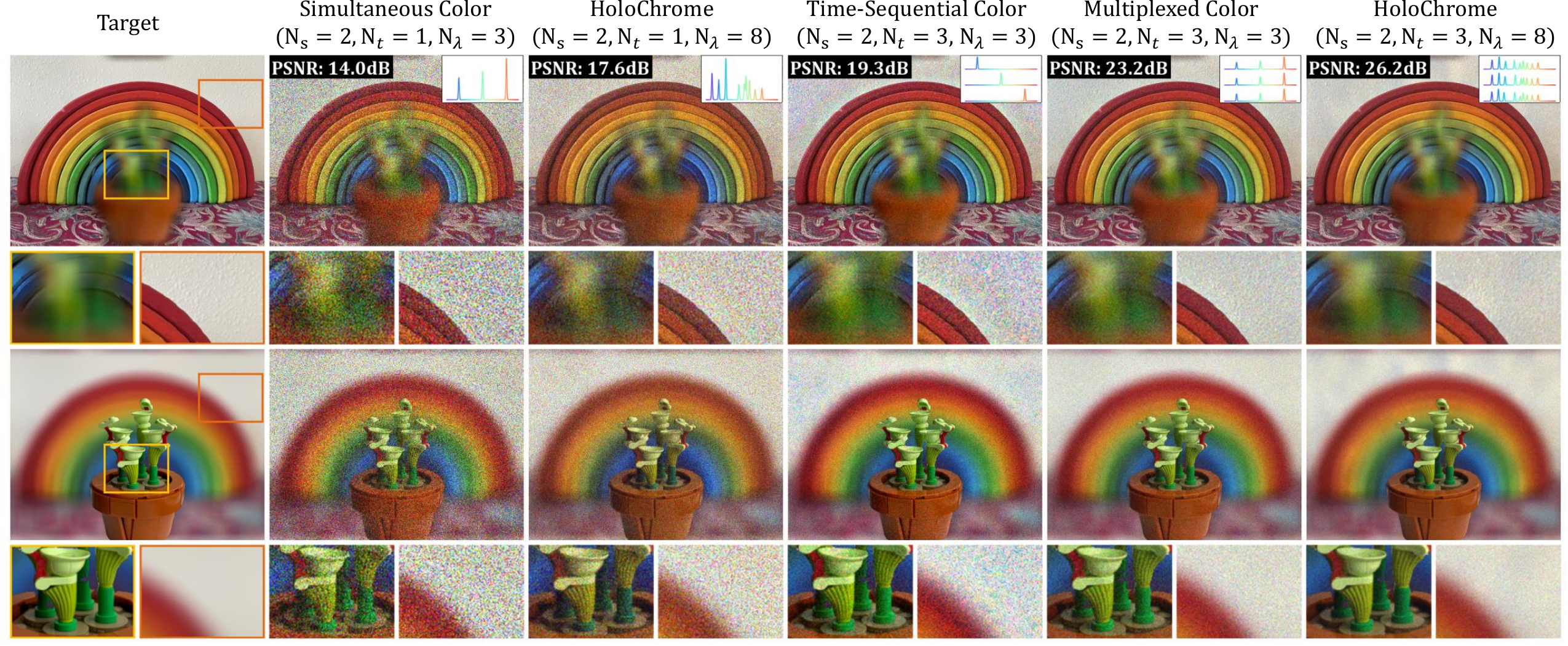}
\caption{
\textbf{Comparison of HoloChrome with conventional holography methods (simulation).}
HoloChrome shows significant performance gains in speckle reduction compared to conventional methods.
The figure compares the results of HoloChrome against three common holographic methods:
Simultaneous Color, Time-Sequential Color, and Multiplexed Color.
The PSNR values indicate that HoloChrome, both in single-frame and three-frame configurations, achieves higher image quality, with the three-frame HoloChrome providing over a 6dB improvement compared to conventional time-sequential color.
Despeckled results are evident in the insets. Illumination spectra are shown in the top right of each column.
\textit{Note: while simultaneous color~\cite{markley2023simultaneous}, time-sequential color~\cite{lee2020wide, curtis2021dcgh, choi2022time}, and multiplexed color~\cite{aksit2023holohdr} techniques have been demonstrated with single SLM architectures, here we use a dual-SLM architecture ($N_s=2$) for fair comparison against HoloChrome.}
}
\label{fig:motivation_1_slm}
\end{figure*}

This paper addresses the above problem by attempting to answer the following question:
How can we optimize spectral illumination for holographic near-eye displays to enhance the perceived image quality?
We develop differentiable Computer-Generated-Hologram (CGH) optimization routines similar to state-of-the-art holography displays, but incorporate arbitrary source spectral profiles and perceptually-motivated color response functions together with hardware constraints on SLM speed and resolution (see Fig.~\ref{fig:concept_figure_forward_model}).
We learn a differentiable hyperspectral hologram model (see Fig.~\ref{fig:learned_holochrome_model}) and show through extensive simulations (Figs.~\ref{fig:motivation_1_slm}-\ref{fig:color_gamut_comparison}) and experiments (Figs.~\ref{fig:teaser_figure}, \ref{fig:experimental_temporal_multiplexing_2d}-\ref{fig:experiment_focalstack_robots}) that polychromatic illumination can significantly reduce speckle noise when compared to time-sequential holographic displays that only illuminate with one narrowband laser source at a time. In particular, we show that polychromatic illumination can reduce speckle noise by as much as 5-6dB in simulation and 3-4dB in experiment, relative to time-sequential RGB color holography architectures.

This paper introduces HoloChrome, a novel polychromatic holographic display system that reduces speckle noise in a manner complementary to existing speckle reduction techniques (e.g. time-multiplexing~\cite{choi2022time}, Multisource~~\cite{kuo2023multisource}, partial coherence \cite{moon2014holographic, peng2021speckle,primerov2019p}). Furthermore, HoloChrome has the added advantage of increasing the color gamut relative to existing three-laser RGB architectures. 
Our proposed system utilizes an ultrafast, wavelength-tuneable laser source to generate a set of independent polychromatic, spatially coherent wavefronts.
The core idea is to incoherently multiplex the images created from these discrete wavelengths, which are all integrated due to the finite response time of the retina. We treat the polychromatic illumination wavelengths $\lambda_i, i\in \{1,...,N_{\lambda}\}$ as trainable parameters that can be optimized on a per-scene basis because the choice of wavelengths has a strong effect on the speckle contrast and color fidelity of displayed images.
When the polychromatic wavelengths are spread far enough from each other along the visible spectrum, they each produce sufficiently decorrelated speckle patterns.
Since the speckle fields produced by each wavelength are mutually incoherent, they do not interfere, and their intensities simply average to reduce speckle contrast.
Furthermore, we utilize the dual-SLM architecture first introduced in \cite{kuo2023multisource} for near-eye displays and later also used in~\cite{chao2024large,kabuli2024practicalhighcontrastholography}. We demonstrate that the dual-SLM architecture helps decorrelate wavelength-dependent speckle fields in a manner similar to Multisource architectures (see supplemental material for detailed discussion), breaking the "memory effect" and enabling effective speckle reduction through incoherent averaging.
Lastly, to accurately model and calibrate our polychromatic holographic system, we develop a hyperspectral propagation model that accounts for the wavelength-dependent Optical Path Differences (OPD) introduced by the optical components. Calibrating the system across all visible wavelengths poses significant challenges due to continuous spectral variations and wavelength-dependent aberrations. We address this by introducing a novel hyperspectral calibration method using a small set of learnable parameters, effectively calibrating the system across all wavelengths with high accuracy and efficiency.
The key innovations of HoloChrome are:

\begin{itemize}
\item \textbf{Polychromatic Illumination:} We use a super-continuum laser to generate a broad spectrum of wavelengths, from which a large number of discrete wavelengths are selectively filtered and multiplexed. This polychromatic illumination enables a significant reduction in speckle noise.
\item \textbf{Dual-SLM Architecture to overcome wavelength correlations:} The spatial separation between two SLMs allows for the decoupling of speckle patterns across different wavelengths. This decoupling enables effective speckle reduction through incoherent averaging of the individual speckle patterns.
\item \textbf{Polychromatic Simulation Framework:} Our framework models key aspects of the optical system, wavelength selection, speckle reduction, and color gamut analysis, as well as perceptual loss. The framework enables the optimization of display performance given a set of hardware constraints.
\item \textbf{Hyperspectral Calibration:} Our novel calibration method calibrates a holographic system for every wavelength in the visible spectrum using just a few learnable parameters.
\item \textbf{Experimental Validation:} We demonstrate a HoloChrome prototype that achieves significant improvements in color reproduction and speckle contrast over conventional holographic displays. 
\end{itemize}

\section{Related Work}
\label{sec:related_work}

\subsection{Accommodation with Incoherent Displays}
The debate on holographic versus incoherent displays centers on the potential of incoherent architectures to address wavefront shaping capabilities, particularly accommodation. Several approaches have been proposed in this realm, including Near-Eye Lightfield Displays (NELD) \cite{lanman2013near}, Split-Lohmann Displays (SLD) \cite{qin2023split}, Pinlight Displays \cite{maimone2014pinlight}, and Multifocal Displays \cite{chang2018towards}. 

Each technology offers distinct advantages but also encounters significant challenges. NELDs, for instance, fall short of achieving diffraction-limited performance comparable to the human eye's maximum acuity. Pinlight Displays face issues with image quality and brightness, while Multifocal Displays struggle with complexity and refresh rates. SLDs provide excellent image quality with accommodation cues but lack a clear pathway to miniaturization. Moreover, manufacturing challenges for foveated near-eye displays, as discussed by \citet{akcsit2019manufacturing}, add further complications to the development of these technologies.
In light of these challenges, holography remains the sole wavefront shaping display technology capable of passing the VTT, while offering a feasible route to compact form factors~\cite{jang2022waveguide,goodman2007speckle}.
ear-eye displays.

\subsection{Challenges in Holographic Displays}
Despite significant advancements, holographic displays still face several challenges that limit their widespread adoption 
\textit{Speckle Noise}
One of the primary challenges in creating high-quality holographic images is speckle noise, which can severely degrade perceived image quality and is the main focus of this paper. Speckle noise is a particularly challenging problem for random phase holograms that produce uniform eyebox intensity and produce the most convincing accommodation cues~\cite{kavakli2023realistic, yang2022diffraction} 

\textit{Etendue}
Another major challenge is the pixel pitch and count of commercially available SLMs, which determines the achievable etendue and constrains the achievable combination of Field-of-View and eyebox size of holographic displays. 
Without access to SLMs with smaller pixel pitch and higher pixel density, etendue expansion techniques \cite{tseng2024neural,monin2022exponentially,chae2022etendue} are necessary to achieve eyebox sizes comparable to conventional 2D near-eye displays. However, these techniques tend to increase speckle noise, further emphasizing the need for advanced speckle reduction methods.
In addition, recent work by Chao et al. \cite{chao2024large} employs a dual-SLM in combination with a multisource approach \cite{kuo2023multisource} to increase etendue while maintaining low speckle-noise levels.

\textit{2D versus 3D Holograms}
Creating 2D holograms is generally an easier task than creating 3D holographic imagery.
In 2D holography, the primary goal is to generate a speckle-free image within a single plane.
This poses a relatively well-conditioned phase-retrieval problem as the number of pixels to be displayed is aligned with the number of SLM pixels. 
In contrast, 3D holograms, designed to provide accurate defocus cues, pose an ill-posed optimization problem.
In this case, we ask the SLM to control a 3D volume with only a 2D modulation.
As a result, the physical system contains insufficient degrees of freedom that results in perceived images with speckle noise.

\subsection{Speckle Reduction}
To mitigate speckle noise in holographic displays, several approaches have been explored, each with its own advantages and limitations:

\textit{Smooth phase holograms:} These holograms have been proposed as a method to eliminate speckle by enforcing a spatially smooth phase distribution at the image plane \cite{maimone2017holographic, shi2021towards}. While effective in reducing speckle, these methods are limited in their ability to generate natural-looking defocus blur and drive accommodation \cite{kavakli2023realistic, yang2022diffraction}. This trade-off between speckle reduction and the ability to create realistic focal cues is a key challenge in holographic display design.

\textit{Random phase holograms:}, These holograms are capable of producing more realistic defocus cues but suffer from inherent speckle noise \cite{yoo2021optimization,schiffers2023stochastic}. Iterative approaches have been employed to shape the phase at the image plane and minimize speckle from a particular viewing angle. However, these methods often require computationally intensive optimization processes and may not generalize well to different viewing positions or display configurations.

\textit{Partially coherent illumination:} These methods use partially coherent sources, such as LEDs \cite{moon2014holographic, peng2021speckle} or superluminescent LEDs \cite{primerov2019p}, and have been explored as a means to reduce speckle. By introducing a degree of incoherence in the illumination, these methods can effectively suppress speckle formation. However, this often comes at the cost of reduced image resolution or depth of field \cite{lee2020light}, limiting the practical applicability of partially coherent illumination in high-quality holographic displays.

\textit{Temporal multiplexing:} These techniques have shown promise in reducing speckle by displaying multiple holograms with unique speckle patterns in rapid succession \cite{lee2020wide, curtis2021dcgh, choi2022time}. By averaging the speckle patterns over time, the perceived speckle contrast can be significantly reduced. However, this approach requires high-speed SLMs and may limit the available bandwidth for other display features, such as color reproduction or dynamic range.

\textit{Polarization multiplexing:} These methods exploit the orthogonal polarization states to generate decorrelated speckle patterns \cite{nam2023depolarized}. By utilizing the two orthogonal polarization states, the methods can effectively reduce speckle noise. However, the number of multiplexed frames is limited to two due to the finite number of polarization states, which may not be sufficient for achieving the desired level of speckle reduction in complex display scenarios.

\begin{figure*}[ht]
\centering
\includegraphics[width=1.0\linewidth]{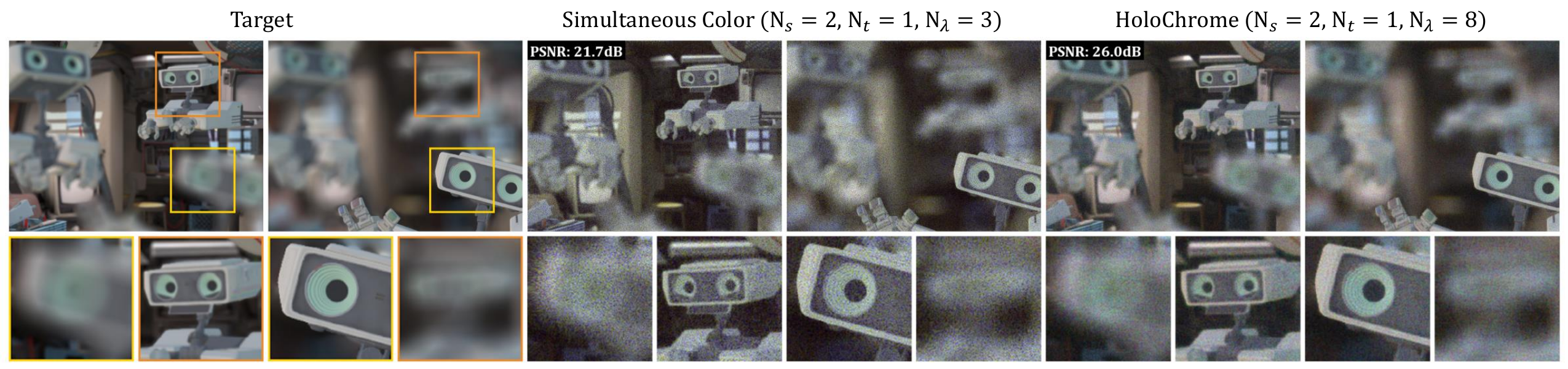}
\caption{
    \textbf{Single frame ($N_t=1$) HoloChrome reduces speckle noise (simulation).}
    ~
    Comparison of image quality with varying wavelengths for a single SLM frame.
    The first column shows the target focal stack.
    The second column shows a simultaneous color~\cite{markley2023simultaneous} result with 3 wavelengths optimized in a single frame, producing a PSNR of 21.7dB.
    The third column shows a HoloChrome result with 8 wavelengths, resulting in a PSNR of 26.0dB.
    The results demonstrate that HoloChrome effectively reduces speckle noise in single frame, with more wavelengths yielding better image quality and more accurate color.
    \textit{Note: while simultaneous color~\cite{markley2023simultaneous}, was demonstrated with a single SLM architecture, here we use a dual-SLM architecture ($N_s=2$) for fair comparison against HoloChrome.}
    }
\label{fig:single_frame_holochrome_works}
\Description{Single Frame Holochrome Works}
\end{figure*}

\textit{Multisource holography:} This method is likely the most promising solution for addressing the speckle problem in holographic displays \cite{kuo2023multisource}. The method reduces speckle noise by employing multiple spatially separated sources, effectively removing noise and enhancing image quality. Multisource holography inherently requires two SLMs because using a single SLM leads to an ill-posed problem where direct image copies created by each source in the grid complicate the deconvolution process.

A notable drawback of the original Multisource method is its reliance on time-sequential display for color reproduction.
While combining Multisource holography together with simultaneous and/or multiplexed color holography methods \cite{markley2023simultaneous, kavakli2023multi} is a logical extension of the original approach, to our knowledge, we are the first to explore the idea of combining multiplexed color with Multisource in simulation (see Fig.~\ref{fig:multisource_vs_holochrome}).

\textit{Wavelength Multiplexing:} This method (a.k.a. HoloChrome) is the focus of this paper, and offers an orthogonal speckle reduction method that leverages wavelength diversity to achieve independent speckle patterns.
By combining the advantages of wavelength multiplexing and a dual-SLM design, HoloChrome aims to achieve high-quality images with significantly reduced speckle noise while maintaining a reasonably large color gamut. 

\textit{Color Reproduction in Holographic Displays}
Recently, researchers have introduced innovative approaches for color reproduction in holographic displays such as simultaneous color \cite{markley2023simultaneous}, which achieves color rendering in a single frame, and multiplexed color~\cite{kavakli2023realistic}, which has demonstrated improved performance for smooth-phase holograms.
However, the validation of both methods with random phase holograms remains untested.

\begin{figure*}[ht]
\centering
\includegraphics[width=1.0\textwidth]{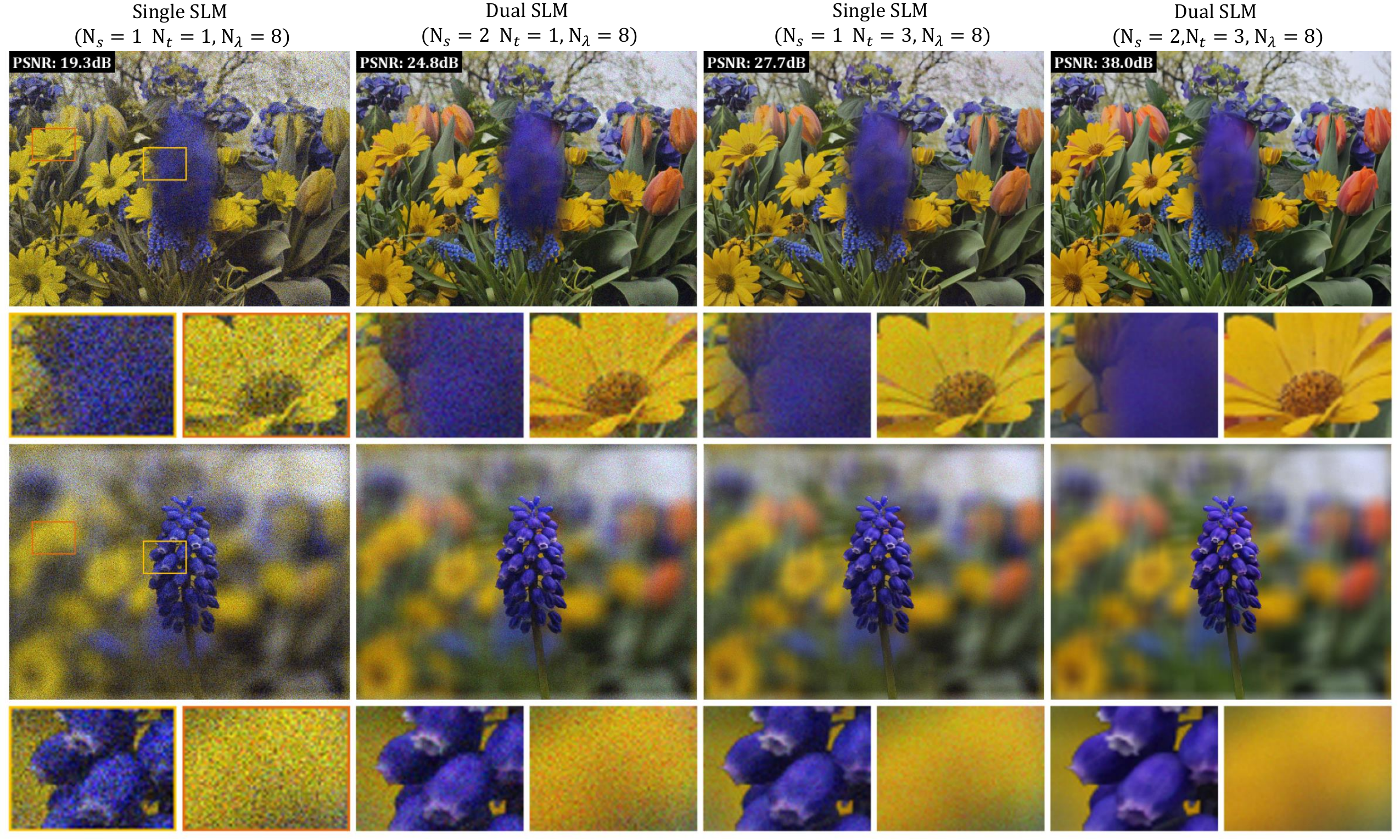}
\caption{
\textbf{Comparison of 1 vs 2 SLMS~(simulation, both phase-only).} 
Introducing a dual-SLM configuration effectively mitigates the wavelength-dependent memory effect.
The figure compares single-SLM and dual-SLM setups in both single-frame and three-frame configurations.
The dual-SLM approach shows a clear reduction in speckle noise, with PSNR values indicating significant performance gains.
In the three-frame configuration, the dual-SLM setup achieves near-perfect image reconstruction.
}
\label{fig:comparison_1_vs2_slms}
\end{figure*}

\section{HoloChrome Framework}
\label{sec:proposed_method}

This section introduces the HoloChrome framework, as depicted in Fig.~\ref{fig:concept_figure_forward_model}, which merges the advancements from \citet{kuo2023multisource} and \citet{kavakli2023multi}.
Designed to enhance holographic imaging, HoloChrome integrates a hyperspectral propagation model with a perceptual eye response function, addressing color losses and reducing speckle noise through wavelength multiplexing and complementing Multisource illumination techniques introduced by \citet{kuo2023multisource}.
Additionally, the framework employs perceptually correct color loss functions and polychromatic illumination for more accurate color representation in holographic displays. 

With HoloChrome, we have developed \textbf{the first wavelength-dependent, differentiable, hyperspectral hologram model that supports camera-in-the-loop calibration}, yielding high-quality experimental results that confirm our theoretical advancements.
The discussion extends to the wavelength continuous forward model, capturing the complete light propagation cycle through dual SLMs, including wavelength-specific aberrations.
Incorporating perceptual color-weighting functions into our optimization routines, we enhance the representation of colors as perceived by human vision via perceptual loss functions.

\subsection{Polychromatic Source Modeling}
\label{sec:hyperspectral_image_formation}
Let $\lambda_i$ denote the $i$-th wavelength selected from the polychromatic source, where $i \in \{1, 2, \dots, N_{\lambda}\}$ and $N_{\lambda}$ is the total number of wavelengths.
The 2D complex field representation of the source-field that illuminates the first SLM with the $i$-th wavelength can be written as:
\begin{equation}
    s(\mathbf{x}, \lambda_i) = e^{j (\mathbf{x}\cdot\mathbf{m}_i)},
    \label{eq:angle_of_incidence_wavelength}
\end{equation}
where $\mathbf{x}$ represents the 2D spatial coordinates on the SLM plane and $\mathbf{m}_i$ is the wavelength-dependent phase slope (in radians per meter) of the $i$-th wavelength at the SLM plane. 
%
%
For an ideal, collimated system there is little chromatic dispersion, and therefore $\mathbf{m}_i$ is typically close to 0.
However, a real, experimental bench-top requires extensive calibration of spatially-varying source aberrations, as detailed later in the paper (see Sec.~\ref{subsec:Wavelength Dependent Source Model}).

\subsection{Polychromatic SLM Response}
\label{sec:LUT}
Each pixel of a SLM modulator has a wavelength-dependent phase-retardation function that maps a grayscale level to the corresponding phase-delay.
Hence, we need to define a wavelength-dependent Look-up-Table $\text{LUT}( g; \lambda ) $, which maps a bit-level value $g$ to the corresponding phase-shift.

Our LUT is an idealized model that doesn't take into account chromatic dispersion in the SLM. This model is accurate for our experimental system as it uses mirrors to shift the optical path (see Sec.~\ref{subsec:Wavelength_slm_model_calibration} for details), but a system using LCoS SLMs would likely require an updated model that incorporates chromatic dispersion.
For more details, we refer to the calibration section found later in the paper in Sec.~\ref{subsec:Wavelength_slm_model_calibration}.

The grayscale value at each spatial location $\mathbf{x}$ on the SLM is denoted as $g(\mathbf{x})$. For simulations, we ignore quantization so that $g(\mathbf{x}) \in \mathbb{R}$, while for our experiments with 4-bit SLMs $g(\mathbf{x}) \in \{0,...,15\}$. The phase-shift $\phi_k$ and diffracted field $p_k$ induced by the SLM at a given wavelength $\lambda_i$ can thus be expressed using the LUT as follows:
\begin{align}
    \phi_k(\mathbf{x}, \lambda_i) &= \text{LUT}(g(\mathbf{x}), \lambda_i), \\
    p_k(\mathbf{x}, \lambda_i) &= e^{i\phi_k(\mathbf{x}, \lambda_i)}
\end{align}
For both simulations and experiments, we model the LUT as a linearly separable function of the grayscale value $g(\mathbf{x})$ and the wavelength $\lambda_i$. This can be expressed as:
\begin{equation}
    \phi_k(\mathbf{x}, \lambda_i) = \alpha(\lambda_i) \cdot g(\mathbf{x})
\end{equation}
where $\alpha(\lambda_i)$ is wavelength dependent scale factor. Our model ignores chromatic dispersion so that the scale factor is proportional to wavelengths and $\alpha(\lambda_i) = k\cdot \lambda_i$, where $k$ is a constant. For our experiments, we learn the values $\alpha(\lambda_i)$ for a set of anchor wavelengths using model training, as described in Section~\ref{sec:calibration_procedure}.

\begin{figure*}[tb]
\centering
\includegraphics[width=1.0\linewidth]{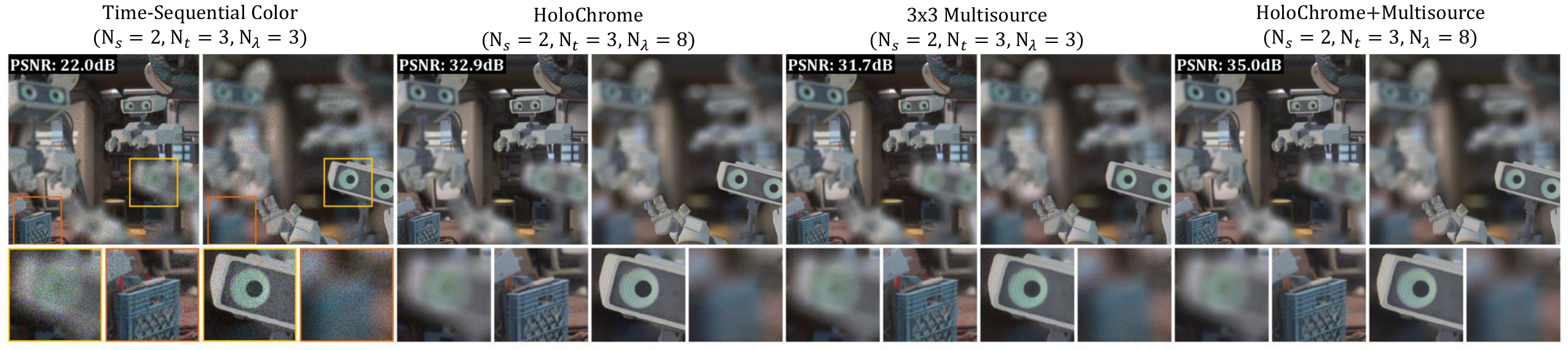}
\caption{
\textbf{Multisource vs HoloChrome~(simulation).}
The figure compares the performance of Color Sequential, HoloChrome, Multisource, and a combination of HoloChrome and Multisource methods for speckle reduction, all using 3 SLM frames to form the full image to ensure a fair comparison.
We also report the mean PSNR for each experiment over the full Focal Stack.
The images demonstrate that both HoloChrome and Multisource methods work reasonably well in reducing speckle.
However, they achieve speckle reduction through independent methods.
The combination of both methods further improves the PSNR, indicating a synergistic effect.
  }
\label{fig:multisource_vs_holochrome}
\Description{Comparison of performance for speckle reduction using different methods.}
\end{figure*}

\subsection{Propagation Model}

We begin by modeling the propagation from the first to the second SLM using the angular spectrum propagation operator (ASM). The propagation over a fixed distance \(\Delta z\) is expressed as:
\begin{equation}
    \mathcal{P}_{\Delta z, \lambda_i} \left( p_1(\mathbf{x};\lambda_i) \cdot s(\mathbf{x}, \lambda_i) \right)
\end{equation}
Here, \(\mathcal{P}_{\Delta z, \lambda}\) denotes the ASM propagation operator, \(\Delta z\) is the fixed distance between the two SLMs, and \(p_1(\mathbf{x};\lambda_i) \cdot s(\mathbf{x}, \lambda_i)\) represents the initial complex field after $\text{SLM}_1$ was applied.
The ASM propagation operator for a given wavelength \(\lambda\), propagation distance \(z\), and a complex field \(f(\mathbf{x})\) is defined as:
\begin{align}
    \mathcal{P}_{z, \lambda}\left\{f(\mathbf{x})\right\} &= \mathcal{F}^{-1} \left\{ \mathcal{F}\{f(\mathbf{x})\} \cdot \mathcal{H}_{z, \lambda}(\mathbf{u}) \right\}, \label{eq:ASM_propagation} \\
    \mathcal{H}_{z, \lambda}(\mathbf{u}) &= 
    \begin{cases} 
        \exp \left( \frac{2 \pi j z}{\lambda} \sqrt{1- \|\lambda \mathbf{u} \|^2} \right), & \text{if } \sqrt{\|\mathbf{u}\|^2} < \frac{1}{\lambda}, \\
        0, & \text{otherwise}.
    \end{cases}
    \label{eq:ASM_kernel}
\end{align}
%
%
Here, \(\mathcal{F}\{\cdot\}\) is the 2D Fourier transform operator and \(\mathbf{u}\) is the 2D coordinate in frequency space~\cite{matsushima2009band}.
The intensity at the sensor plane located at propagation distance \(z\) from the second SLM can be expressed as:
\begin{equation}
    I^{m}(\mathbf{x},z; \lambda_i) = \left| \mathcal{P}_{z, \lambda_i} \left( \mathcal{P}_{\Delta z, \lambda_i} \left( p_1(\mathbf{x}, \lambda_i) \cdot s(\mathbf{x},\lambda_i) \right) \cdot p_{2}(\mathbf{x},\lambda_i) \right) \right|^2 .
    \label{eq:hyperspectral_forward_model}
\end{equation}

We now introduce the laser amplitude \(a_i\), which controls the intensity for laser wavelength \(\lambda_i\).
When computing the hologram, $a_i$ will be optimized, as the optimal weighting between the wavelengths is not known as a priori.
The overall signal intensity measured at a specific wavelength for a given laser power is defined as:
\begin{equation}
    I(\mathbf{x,z};\lambda_i, a_i, g) = a_i \cdot I^{m}(\mathbf{x},z;\lambda_i, g),
\end{equation}
where we have explicitly denoted dependence on the two SLM patterns used in our HoloChrome setups \(g = (g_1(\mathbf{x}), g_2(\mathbf{x}))\).

\subsection{Aperture Effects and Aberration Considerations}
\label{subsec:modeling_spatial_varying_aberrations}

In an ideal optical setup, aberrations are typically negligible, and the dependence on the wavelength is minimal. However, experimental setups often exhibit significant aberrations which necessitate careful calibration (refer to Sec.~\ref{subsec:modeling_spatial_varying_aberrations} for further details).

In holographic and computational imaging prototypes, apertures are used strategically to manage higher-order aberrations and block unwanted DC components. The effect of each wavelength passing through the aperture differs due to variations in the spatial frequency cut-off. This phenomenon needs to be precisely modeled, as depicted in Fig.~\ref{fig:concept_figure_forward_model}.

We can model the wavelength-dependent spatial frequency cut-off by incorporating an aperture into the angular spectrum propagation operator (ASM). The modified propagation operator \(\mathcal{P}_z\) that includes the aperture is defined as follows:
\begin{equation}
        \mathcal{P}_{z, \lambda}\left\{f(\mathbf{x})\right\} = \mathcal{F}^{-1} \left\{ \mathcal{F}\{f(\mathbf{x})\} \odot \mathcal{H}_{z, \lambda}(\mathbf{u}) \odot \mathcal{A}(\mathbf{u}, \lambda) \right\},
        \label{eq:ASM_propagation_with_aperture}
\end{equation}
where $\mathcal{A}(\mathbf{u}, \lambda)$ ensures that the propagation model accounts for the impact of the physical aperture on different wavelengths.
%
In the general case, $\mathcal{A}(\mathbf{u}, \lambda)$ denotes a complex pupil function, which encapsulates both the amplitude transmission and the optical path difference (OPD) effects:
\begin{align}
    \mathcal{A}(\mathbf{u}, \lambda) = \mathcal{T}(\mathbf{u}, \lambda) \cdot e^{i\cdot \frac{2\pi}{\lambda}\text{OPD}(\mathbf{u},\lambda)},
    \label{eq:OPD}
\end{align}
where \( \mathcal{T}(\mathbf{u}, \lambda) \) represents the amplitude function and \( \text{OPD}(\mathbf{u}, \lambda) \) is the optical path difference across the aperture for different wavelengths that can model aberrations. Our simulations assume an ideal, constant \( \mathcal{A}(\mathbf{u}, \lambda) = 1 \) without any aberrations. In a real system, the exact form of the aberration function is not known a priori and is learned via the training procedure outlined in Sec.~\ref{sec:calibration_procedure}. Together, the source term \( s(\mathbf{x}, \lambda) \) and aperture \( \mathcal{A}(\mathbf{u}, \lambda) = 1 \), represent the main learnable parameters of our hyperspectral calibration procedure.

\begin{figure*}[ht]
\centering
\includegraphics[width=1.0\textwidth]{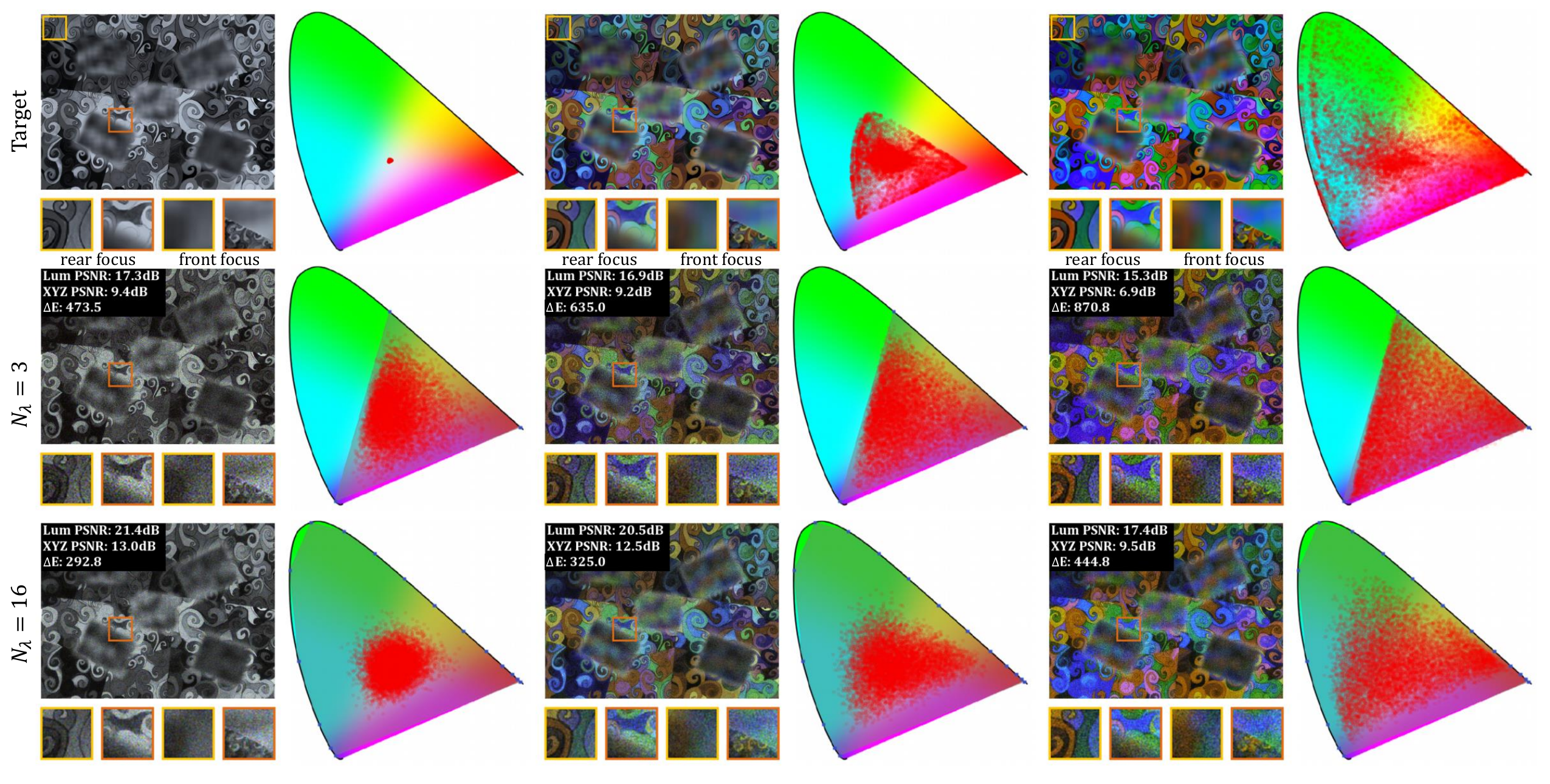}
 \caption{
    \textbf{HoloChrome PSNR versus color gamut (simulation).}
    All examples in the figure use a single frame dual SLM setup ($N_s=2, N_t=1$).
    The figure illustrates the tradeoff between wide color gamut and speckle reduction offered by HoloChrome.
    The first row shows the targets with an increasingly wide color gamut, from left to right. 
    The second and third rows show the reconstructed images for \(N_\lambda = 3\) and \(N_\lambda = 16\) wavelengths, respectively, with varying PSNR values for luminance (Lum PSNR) and XYZ color space (XYZ PSNR), along with color difference (\(\Delta E\)).
    %
    %
    The results highlight the balance between achieving a wide color gamut and reducing speckle noise, with higher numbers of wavelengths providing better speckle reduction at the cost of gamut size.
    }
    \label{fig:color_gamut_comparison}
\end{figure*}

\subsection{From Polychromatic Field to Perceived Color}
To accurately represent the perceived colors of a polychromatic hologram, it is crucial to consider the human visual system's response to different wavelengths.
To achieve this, we first compute the Long, Medium, and Short (LMS) response functions of the human eye \cite{stockman2000spectral}, which weigh the contribution of each wavelength based on the sensitivity of the corresponding cone cells.
The LMS response function $LMS^c(\lambda_i)$ for a channel $c \in [l, m, s]$ is formulated as:
\begin{equation}
LMS^c(\mathbf{x},z;\lambda_i, a_i,g) = \int I(\mathbf{x},z;\lambda_i, a_i,g) \cdot W(\lambda_i)^c , d\lambda_i,
\label{eq:lms_response}
\end{equation}
where $W(\lambda_i)^c$ is the LMS weighting function for channel $c$, and we define the three channel model output as $\mathbf{LMS}(\mathbf{x},z;\lambda_i, a_i,g) \in \mathbb{R}^3$.
This equation integrates over the product of the hologram intensity and the spectral weighting functions to obtain the LMS response for each channel.
However, the LMS response does not provide an ideal colorspace to measure perceptual similarities for human vision \cite{fairchild2013color}.
To improve perceptual accuracy, we convert the estimated LMS response into either sRGB (for narrow-gamut targets) or CIE XYZ (for wide-gamut targets) color space.
\subsection{Incorporating Perceptual Colorspaces}
We define the image in the target color space as \(\mathbf{T}(\mathbf{x},z) \in \mathbb{R}^3\) at propagation distance \(z\).
Each pixel location $(\mathbf{x},z)$ in the target contains a 3-dimensional value defined in a known color space (e.g., as LMS, sRGB, CIEXYZ, CIELUV).
To compute a loss between the model and target, we need to map the model output from LMS color space (Eq.~\ref{eq:lms_response}) into the target color space using a differentiable color transformation.
We denote the 3 color model output $\mathbf{O} \in \mathbb{R}^3$ in the target color space as:
\begin{equation}
    \mathbf{O}(\mathbf{x}, z; \lambda_i, a_i, g) = M\left[\mathbf{LMS}(\mathbf{x},z; \lambda_i, a_i, g)\right],
\end{equation}
where \(M\left[\cdot\right]\) represents the differentiable color transformation from LMS to target colorspaces (e.g., LMS to sRGB).

\subsection{Final Loss Function}

We first define a discretized set of 2D spatial coordinates $\mathbb{X} \in \{(x,y) | x \in [1,..,N_x], y \in [1,..,N_y]\}$ and propagation distances $\mathbb{Z} \in \mathbb{R}^{N_z}$. 
We then define vector valued representations of the model parameters: each of the $N_s$ SLM patterns for each of the $N_t$ time frames $\mathbf{g} \in \mathbb{R}^{N_x\cdot N_y\cdot N_s \cdot N_t} $, wavelength values $\mathbf{l} \in \mathbb{R}^{N_{\lambda}}$, and wavelength amplitudes $\mathbf{a} \in \mathbb{R}^{N_{\lambda}}$. 
We then define our loss as the \(l_2\) MSE-loss between the target and model output, formulating the optimization problem for the SLM pattern as:
\begin{align}
    \hat{\mathbf{s}}, \hat{\mathbf{a}} = \operatorname*{argmin}_{\mathbf{s},\mathbf{a}} \sum_{\mathbf{x} \in \mathbb{X}} \sum_{z \in \mathbb{Z}} \| \mathbf{O}(\mathbf{x},z;\mathbf{s}, \mathbf{a}, \mathbf{l}) -  \mathbf{T}(\mathbf{x},z) \|_2^2,
    \label{eq:optimization_problem_hyperspectral}
\end{align}
In a variant of this loss function, we also optimize the wavelengths $\mathbf{l}$ in addition to the laser amplitude of each wavelength $\mathbf{a}$, as
\begin{align}
    \hat{\mathbf{s}}, \hat{\mathbf{a}}, \hat{\mathbf{l}} = \operatorname*{argmin}_{\mathbf{s},\mathbf{a},\mathbf{l}} \sum_{\mathbf{x} \in \mathbb{X}} \sum_{z \in \mathbb{Z}} \| \mathbf{O}(\mathbf{x},z;\mathbf{s}, \mathbf{a}, \mathbf{l}) -  \mathbf{T}(\mathbf{x},z) \|_2^2,
    \label{eq:optimization_problem_hyperspectral_with_wavelengths}
\end{align}
Note that while our experiments utilize a tunable supercontinuum laser source allowing for an arbitrary choice of wavelengths, the HoloChrome framework is not limited to using laser sources.
If a fixed array of discrete wavelength primaries is used, wavelength optimization during operation is not possible.
However, our simulation experiments show that even with a predetermined set of wavelengths, significant improvements in speckle reduction and color reproduction can still be achieved compared to conventional 3-primary methods. Using fixed discrete wavelengths simplifies the system and enhances practicality for real-world applications.

We found that the choice of color space used inside the loss-function can have quite a significant influence on speckle. For instance, performing hologram optimization in the sRGB space places a higher emphasis on a narrow gamut of color values, while the CIEXYZ space can enforce accurate color representation over a wider gamut.

For all of our color focal stack simulations (Figs.~\ref{fig:motivation_1_slm}-~\ref{fig:multisource_vs_holochrome}), we used an sRGB loss, which we found reduces the speckle and produces the best PSNR. However, we also performed color gamut experiments (see Fig.~\ref{fig:color_gamut_comparison}) using the CIEXYZ loss which produces higher-quality color reproduction, at the cost of some loss in speckle reduction.
We posit that investigating the trade-offs associated with using different perceptual color spaces during optimization is a compelling area of research for holographic displays.
Although this paper does not delve deeply into this subject, we encourage future research to explore these trade-offs in greater detail.

\begin{figure*}[htb]
\centering
\includegraphics[width=1.0\linewidth]{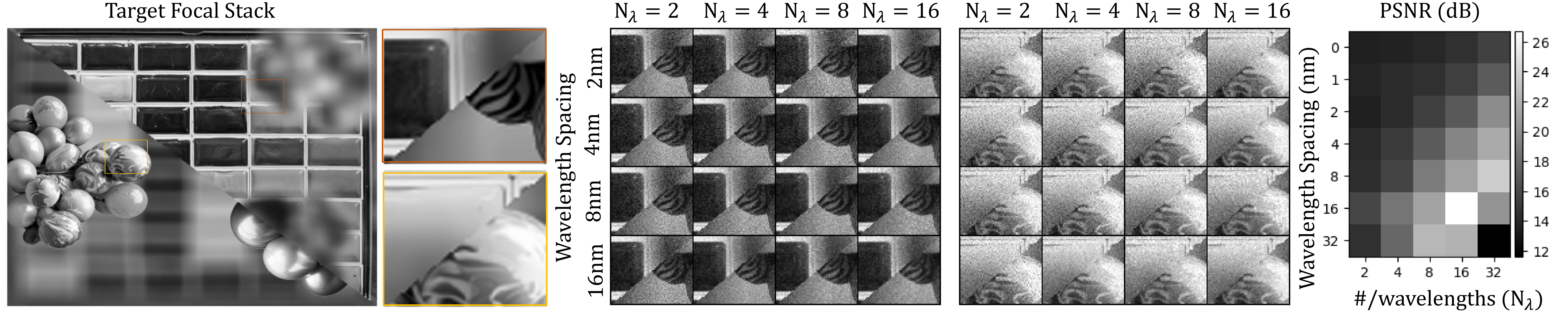}
\caption{
\textbf{Ablation study on spacing and number of wavelengths (simulation).}
All examples in the figure use a dual SLM setup ($N_s=2$).
The figure shows the effect of varying bandwidth and number of wavelengths on image quality.
The leftmost panel displays the target focal stack.
The middle panels illustrate the reconstructed images at different wavelength spacings (2nm, 4nm, 8nm, 16nm) and the number of wavelengths ($N_{\lambda}= 2, 4, 8, 16$), demonstrating the visual improvement in speckle reduction as both parameters increase.
The rightmost panel presents a heatmap of PSNR values, with darker shades indicating lower PSNR and lighter shades indicating higher PSNR.
For this study, optimal performance is achieved for $N_{\lambda}=16$ over $240nm$ bandwidth. 
\textit{Note: In this simulation, wavelengths are evenly spaced across the visible spectrum while for all other simulations in the paper, wavelengths are optimized individually for each target.   }
}
\label{fig:ablation_study_bandwidth}
\Description{Effect of varying bandwidth and number of wavelengths on image quality.}
\end{figure*}

\section{Simulated Image Quality}
\label{sec:simulation_results}

The use of multiple, mutually incoherent wavelengths in our system enables effective speckle reduction through incoherent averaging.
When the intensities of the wavelength-dependent speckle patterns are summed together, the resulting speckle contrast is reduced. 
In this section, we present a set of simulations to analyze speckle reduction performance of HoloChrome systems.  

\textit{Simulation Details:} 
In this section and the next, we introduce a set of simulations to study HoloChrome image quality in terms of speckle reduction and color fidelity.  
Our simulations assume an idealized SLM model without cross-talk, quantization noise, or chromatic dispersion. All simulations in the paper use the following parameters: continuous phase SLMs with $3\pi$ phase range at $550nm$ (and no dispersion), SLM pixel pitch is $8\mu m$, SLM resolution is 512x384 pixels upsampled 2x into 1024x768 pixels with $4\mu m$ pitch. All simulation results are optimized using focal stack targets.
We simulate natural defocus cues for target focal stacks generated from 2D image stacks that incorporate realistic blur effects. 
This blur is modeled based on incoherent illumination, with the kernel size determined by the SLM's maximum diffraction angle.
In some cases, we use 2D image content to generate a focal stack by synthesizing the correct out-of-focus blur at each depth plane, and in all cases, we initialize with a random phase to help ensure uniform eyebox intensity. Propagation between SLMs is $\Delta z = 2mm$, and the focal stack is generated using 4-5 focal planes uniformly spaced from $z = 9mm-11mm$.

\subsection{HoloChrome vs Conventional Holography}

%
\label{section:holochrome_vs_conventional_holography}
We compare HoloChrome against three conventional methods for creating color holograms:
\begin{itemize}
\item \textit{Simultaneous Color}: Diffracted fields from three wavelengths are jointly optimized within a single time-frame \cite{markley2023simultaneous}.
\item \textit{Time-Sequential Color}: The most common architecture, where RGB channels are independently optimized and sequentially displayed with one wavelength at a time \cite{peng2020neural,chakravarthula2019wirtinger}.
\item \textit{Multiplexed Color}: Both color and time are jointly optimized and multiple wavelengths are turned on simultaneously~\cite{kavakli2023multi}.
\end{itemize}

We note that, while both simultaneous and multiplexed color techniques have been demonstrated previously using smooth phase, our paper is the first to evaluate these techniques for random-phase holograms.  
In Fig.~\ref{fig:motivation_1_slm}, we compare HoloChrome against these three methods.
Note that while simultaneous color, time-sequential color, and multiplexed color techniques have been demonstrated previously with single SLM architectures, here we use a dual-SLM architecture ($N_s=2$) for fair comparison against HoloChrome.
We evaluate both single-frame and three-frame reconstruction algorithms.
For the single frame case, HoloChrome with $N_{\lambda}=8$ performs 3.6dB better than using only three colors~\cite{markley2023simultaneous}.

When we extend the time budget to three frames, HoloChrome achieves a 6.7dB PSNR boost over simultaneous color~\cite{markley2023simultaneous}, and 3.5dB boost over multiplexed RGB-color~\cite{kavakli2023multi}.

Overall, HoloChrome demonstrates significant performance gains over conventional methods, particularly in reducing speckle, thereby enhancing image quality.

\subsection{HoloChrome versus Simultaneous Color}
\label{sec:improvement_simultaneous_color}

Our next comparison further illustrates the benefit of multiplexing multiple wavelengths within a single SLM frame. Figure~\ref{fig:single_frame_holochrome_works} demonstrates that HoloChrome with $N_{\lambda}=16$ produces 4.3dB improvement in speckle noise reduction relative to simultaneous color~\cite{markley2023simultaneous}, which only multiplexes 3 wavelengths into a single frame. These results demonstrate that HoloChrome effectively incorporates more wavelengths with greater speckle diversity, yielding better image quality with less speckle contrast.

\subsection{Improvements using 1 SLM vs 2 SLM}
\label{sec:improvement_1slm_vs_2slm}
In Sec.~\ref{section:holochrome_vs_conventional_holography}, we demonstrated that HoloChrome provides significant performance improvements using a single SLM.
However, to further mitigate the wavelength-dependent memory effect, we explored the use of dual SLMs, as detailed in the methods section.
Figure~\ref{fig:comparison_1_vs2_slms} compares the performance of using one SLM versus two SLMs in both single-frame and time-multiplexed configurations.
The dual-SLM setup effectively breaks the wavelength-dependent memory effect, resulting in a noticeable reduction in speckle noise.

For the single-frame configuration, the dual-SLM approach significantly enhances image quality by 5.3dB compared to the single-SLM case.
%
%
When extending the time budget to three frames, the dual-SLM configuration achieves near-perfect image reconstruction, as indicated by the substantial PSNR gain and the despeckled images in the insets.
This improvement underscores the advantage of utilizing multiple SLMs for holographic displays, leading to superior image quality and reduced speckle noise.
This is in line with the evidence shown from ~\cite{kuo2023multisource}.

\subsection{Comparison with Multisource}

HoloChrome is a speckle reduction method that leverages wavelength diversity to achieve uncorrelated speckle patterns in a manner similar, yet orthogonal to the angular diversity approach used in Multisource Holography.

To demonstrate this orthogonality, we compare the two methods in Fig.~\ref{fig:multisource_vs_holochrome}.
This comprehensive comparison highlights the effectiveness of HoloChrome, Multisource, and their combination in speckle reduction. Both 9-wavelength holochrome and 3x3 multisource produce similar performance benefits over conventional, single-source time-sequential color, while combining the two together produces an additional 2-3db performance boost.

\begin{figure*}[tb]
    \centering
    \includegraphics[width=1.0\linewidth]{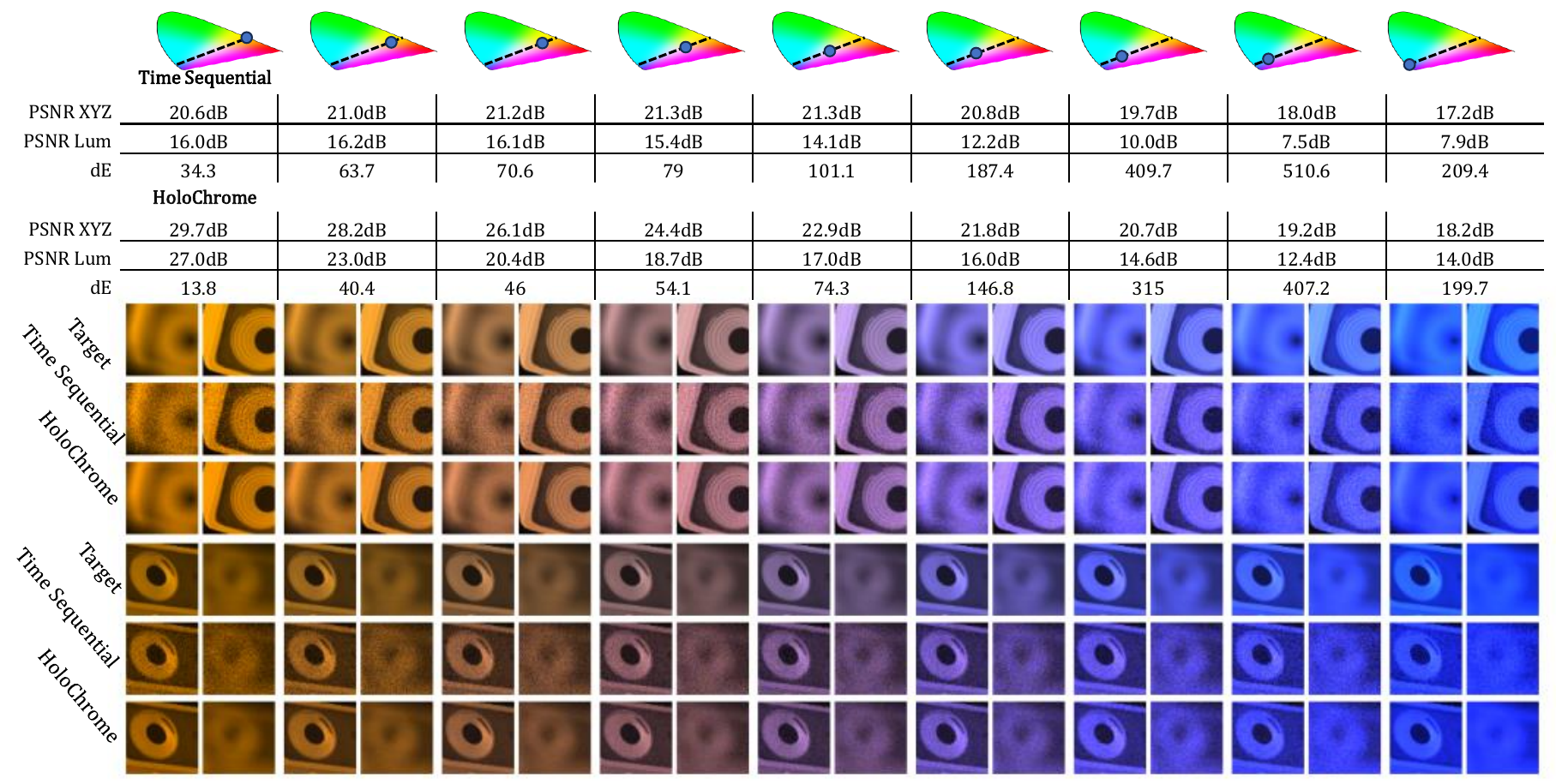}
    \caption{
\textbf{Wavelength multiplexing reduces speckle (simulation).}
All examples in the figure use a two frame, two wavelength, dual SLM setup ($N_s=2, N_t=2, N_{\lambda}=2$).
This ablation study investigates our premise that wavelength multiplexing allows for greater speckle reduction compared to a time-sequential display.
For this figure, we define a mono-color focal stack target that varies on a linear curve in xyY-chromaticity space, where we've rendered out the corresponding color in the image.
We then allow exactly 2 frames for temporal-multiplexing.
For a conventional, color-time sequential, the sensor integrates over 2 frames with 1 wavelength each.
For the proposed HoloChrome system, we allow both wavelengths to be turned on for each of the 2 frames.
For evaluation, we compute PSNR in XYZ and Lumanicity (from CIE LUV) and the averaged color difference (CIE2000 $\Delta$E).
Our proposed holochrome approach consistently outperforms the time-sequential approach.
}
\label{fig:ablaton_wavelength_reduces_speckle}
\Description{Ablation study on wavelength selection and speckle reduction.}
\end{figure*}

\subsection{Color Gamut Analysis}
\label{color_gamut_analysis}

HoloChrome introduces a fundamental trade-off between speckle reduction and color gamut of displayed images. On the one hand, the choice of more than 3 wavelengths introduces the possibility to display images that span a much larger color gamut. However, saturated colors near the border of the gamut cannot be reproduced with the same amount of speckle reduction as low-saturation colors in the center of the gamut. Low-saturation colors can be reproduced with the least amount of speckle noise since independent speckle patterns produced by each source wavelength can all contribute to reproducing colors in the middle of the gamut.
This is not the case for highly saturated colors, where, for example, the speckle of red wavelengths cannot be used to help reduce speckle of blue pixels.

In Fig.~\ref{fig:color_gamut_comparison}, we analyze how the image content, particularly the range of desired colors or gamut in a scene, affects different holographic display prototypes.
To perform this analysis, we first create an artificial 3D target defined in XYZ color space that encompasses colors across the entire human-visible spectrum.
We then transform the target into luminance, saturation, and hue (using LCHuv representation) so that we can effectively change the "color gamut" of our scene by scaling the saturation channel.
The top row of Fig.~\ref{fig:color_gamut_comparison} shows an image of targets with varying saturation, together with  
their respective xy-chromaticity plots on top of the horseshoe gamut representing human-viewable colors.
In the second and third rows, We use these targets to optimize two HoloChrome systems with $N_{\lambda}=3$ and $N_{\lambda}=16$, respectively.
For all simulation experiments, we perform all optimization in one-time frame ($N_{t}=1$), acknowledging that this will naturally lead to noisy images but believing it provides the fairest comparison. We note the following key observations:

\textit{Speckle Reduction:}
As expected, the zero-saturation target results in the lowest amount of speckle noise since all wavelengths can contribute to speckle reduction when reproducing low-saturation colors.

In addition, $N_{\lambda}=16$ always reduces speckle more relative to $N_{\lambda}=3$ regardless of the target color gamut, but the increase in luminance PSNR is $4.1dB$ for zero saturation versus 2.1dB for full saturation.

\textit{Color Accuracy:}
%
The gamut projections show that the 3-primary system cannot create colors outside its spanned gamut. 
In contrast, the 16-wavelength system can create colors that a conventional system cannot.
For both cases, the color error (measured by CIE2000 $\Delta E$) decreases as the saturation decreases and the target gamut area is reduced.
However, the system with $N_{\lambda}=16$ produces a more accurate perceptual color reproduction for all saturation levels, with a difference in color error that is smaller by $\Delta E = 473.5-292.8 = 180.7$ for no saturation versus an even larger reduction in color error of $\Delta E = 870.8-444.8 = 426.0$ for full saturation.

\textit{Gamut Trade-off:}
The ability to achieve a wider color gamut is particularly beneficial for reproducing vibrant and varied colors, though it comes with the trade-off of increased computational complexity since the speckle fields for each wavelength must be computed and integrated using Eq.~\ref{eq:lms_response} inside the optimization loop.
Furthermore, creating random phase holograms that span the full gamut is challenging because creating colors that lie purely on the horseshoe boundary can only be reproduced with a single wavelength.
This is evident in the "wide gamut results" on the far right, where the $N_{\lambda}=16$ creates a wider gamut than $N_{\lambda}=3$, the density of reproduced colors decreases significantly near the horseshoe border.

\subsection{Ablation Study on Wavelength Selection}
In Fig.~\ref{fig:ablation_study_bandwidth}, we conduct an ablation study focusing on the relationship between source illumination spectrum and speckle reduction for HoloChrome. 
Unlike all other examples in the paper where we optimize the wavelengths on a per-target basis, \textit{in this study all wavelengths are chosen to be fixed and evenly spaced} in the center of the visible spectrum. While we have found that wavelength optimization can almost always improve PSNR compared to fixed, uniformly spaced wavelengths spread over the visible spectrum, the improvement is largely content-dependent and never greater than around 1-2dB.

The study provides insight into the optimal choice for source illumination spectrum with respect to speckle noise reduction.               
As in all other examples, the magnitude for each wavelength is optimized on a per-target basis. Because we focus on speckle reduction performance here, we ignore color and use a monochromatic target and sensor model (flat sensor response from $400nm-700nm$) instead of color targets, LMS response curves, and a perceptual color loss. 

The heatmap in Fig.~\ref{fig:ablation_study_bandwidth} reveals a general trend that increasing the number of wavelengths enhances the ability to reduce speckle, as it introduces more diversity in the wavelengths used. Similarly, for a fixed number of wavelengths, performance tends to increase as the wavelength spacing is increased. For this example, the optimal performance is attained with $N_{\lambda}=16$ wavelengths and a spacing of $16nm$, with a bandwidth of $240nm$ that covers most of the visible spectrum. However, increasing either the number of wavelengths to $N_{\lambda}=32$ or the spacing to $32nm$ causes some wavelengths to fall outside the visible spectrum, mitigating the effects of wavelength diversity and increasing speckle noise.

\subsection{Ablation Study on Wavelength Multiplexing}

In Fig.~\ref{fig:ablaton_wavelength_reduces_speckle}, we use a simple example to investigate our core premise that wavelength multiplexing enables greater speckle reduction compared to a time-sequential display where only one wavelength is on at a given time.
For this figure, we define a mono-color (\textit{not monochromatic}) focal stack target that varies on a linear curve in xyY-chromaticity space, where we've rendered out the corresponding color in the image.
We then allow exactly 2 frames for temporal multiplexing.
For a time-sequential color approach, the sensor integrates over 2 frames with 1 wavelength on at a time: for each time frame, the wavelength magnitude is optimized for a single wavelength while the other wavelength amplitude is fixed at zero.
For the proposed HoloChrome system, the sensor integrates over 2 frames, but now for each time frame the amplitude of both wavelengths are optimized to the target. As a result, the final output image integrates 4 mutually incoherent speckle fields instead of just two for time-sequential.

For evaluation, we compute PSNR in XYZ and luminance (from CIE LUV) and the averaged color difference (CIE2000 $\Delta$E).
Our proposed HoloChrome approach consistently outperforms the time-sequential approach, providing better speckle reduction and maintaining high image quality, confirming our underlying premise that color multiplexing can be leveraged to improve imaging performance.

\begin{figure}[tb]
    \centering
    \includegraphics[width=\linewidth]{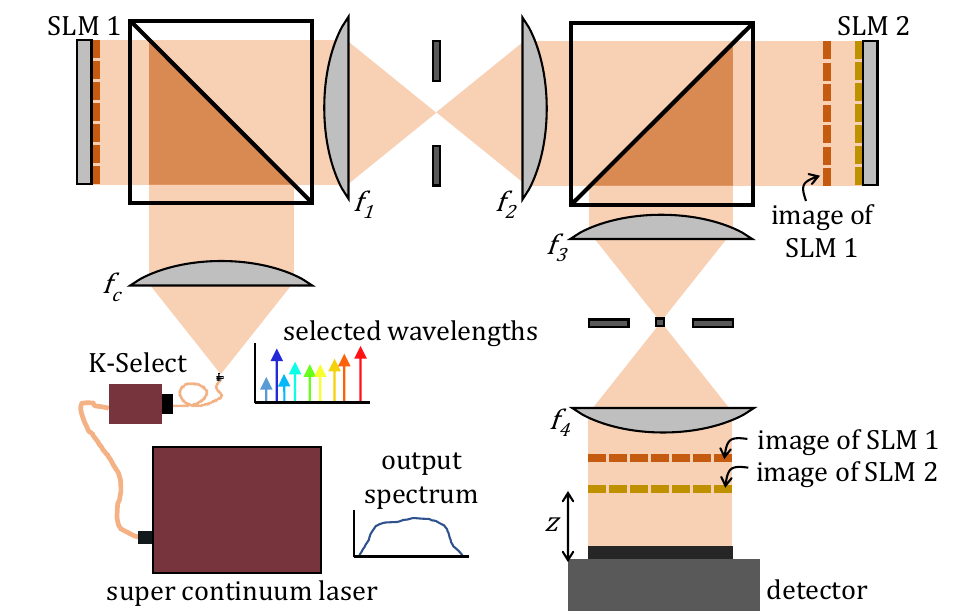}
    \caption{
       \textbf{Schematic of experimental setup.}
       Our benchtop prototype uses two SLMs with a 4$f$ system in between.
       A second 4$f$ relays both SLMs to the correct positions in front of a bare sensor, which is mounted on a linear motion stage.
       Irises in the Fourier planes remove higher orders from the SLMs, and an obscuration in the second Fourier plane blocks the D.C. component.
        A NKT super continuum laser is used for illumination, allowing center wavelength to be tuned anywhere in the visible spectrum.
}
    \label{fig:setup_detailed_sketch}
\end{figure}

\begin{figure*}[tb]
\centering
\includegraphics[width=1.0\linewidth]{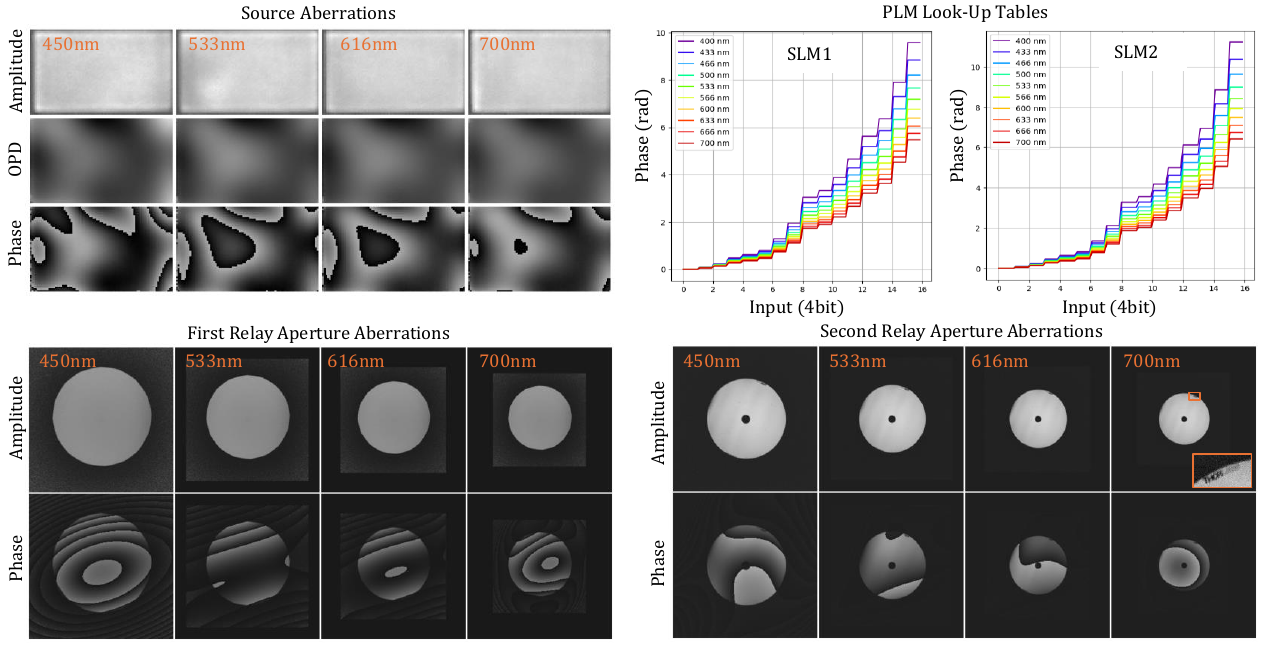}
\caption{
\textbf{Learned hyperspectral HoloChrome model.}
The top left section illustrates 4 of the 10 anchor wavelengths utilized in the model.
Our method independently learns the amplitude and Optical Path Difference (OPD); for enhanced visualization, the wrapped phase of the source field is also presented.
The top-right graph displays the learned 4-bit Look-Up Table (LUT) for both SLM$_1$ and SLM$_2$, covering wavelengths from 400 to 700 nm.
The bottom row depicts the learned amplitude and the corresponding Zernike aberrations for four selected wavelengths for both relays.
The second relay incorporates a DC-filter term.
Notice the quality of the reconstructed aperture term.
The DC-filter used is manufactured by Thorlabs, with a small 'Thorlabs' logo etched on the glass.
Despite not imaging the aperture plane, our calibration procedure is robust enough to reconstruct fine details in the Fourier Domain even though we capture images only in the spatial domain.
}
\label{fig:learned_holochrome_model}
\Description{Learned Holochrome Model}
\end{figure*}

\section{Experimental System Calibration}
\label{sec:experiment_calibration}

In section~\ref{sec:simulation_results}, we demonstrated through simulations that HoloChrome is a promising technique for reducing speckle in near-eye holographic displays.
However, in practice, achieving high-quality experimental results necessitates precise knowledge and characterization of the system parameters, such as the source amplitude and phase, SLM look-up table, the relative positioning of the two SLMs, and aperture aberration functions.
To calibrate our experimental system, we adapt the approaches of \citet{peng2020neural} and \citet{Chakravarthula2020LearnedDisplays} to HoloChrome by designing a physics-inspired forward model where unknown parameters are learned from a dataset of experimentally captured SLM-image pairs.

To the best of our knowledge, all published physics-inspired calibration methods work only for monochromatic illumination.
Therefore, to display RGB images, typically three independent models are learned.
In most cases, a monochromatic image for each color channel is then optimized with the respective model.
However, some approaches, such as~\cite{markley2023simultaneous}, introduce a color-mixing matrix to account for spectral overlap, e.g., the Bayer pattern in an RGB camera.
Similarly, our approach takes into account the physical response of the human eye to ensure correct color reproduction.

One possible approach to implement an experimental HoloChrome model is to optimize an independent model for each wavelength in the visible spectrum.
However, this requires an immense amount of captured data, training time, and storage capacity, with each model still likely to overfit the data.
Given the smooth behavior of physical optics, especially concerning wavelength dependency, we instead directly model the wavelength dependency of each component in one single hyperspectral holographic system.
This approach has several explicit benefits: it significantly reduces the number of learned parameters, making our calibration procedure more robust to local minima, and allows for quicker convergence to a global solution due to lightweight parameterization.
An overview of the learned parameters in our model are shown in Fig.~\ref{fig:learned_holochrome_model}.

\subsection{Hyperspectral Model with Learnable Parameters} \label{sec:learnable_model}

\paragraph{Hyperspectral Source Model}
\label{subsec:Wavelength Dependent Source Model}
In Sec.~\ref{sec:hyperspectral_image_formation}, we mathematically described the polychromatic source $s(\mathbf{x},\lambda_i)$ which is used as a trainable parameter in our HoloChrome models.
A straightforward approach to model the hyperspectral source is to learn a 2D complex field at $N_s$ anchor wavelengths and then perform a 1-D linear interpolation for intermediate wavelengths during evaluation.
However, we found that in order to avoid issues with phase wrapping, it was necessary to learn Optical Path Differences (OPD) directly at these anchor wavelengths instead of phase.
We found that the wavelength-dependent OPD is extremely smooth, making it a suitable representation of our source field with a small number of parameters.
In our implementation, we learn the amplitude and OPD of the source field independently for each anchor wavelength.
During the forward pass, we perform 1D linear interpolation in the wavelength dimension to compute the complex field for arbitrary wavelengths.

\begin{figure*}[tb]
\centering
\includegraphics[width=1.0\textwidth]{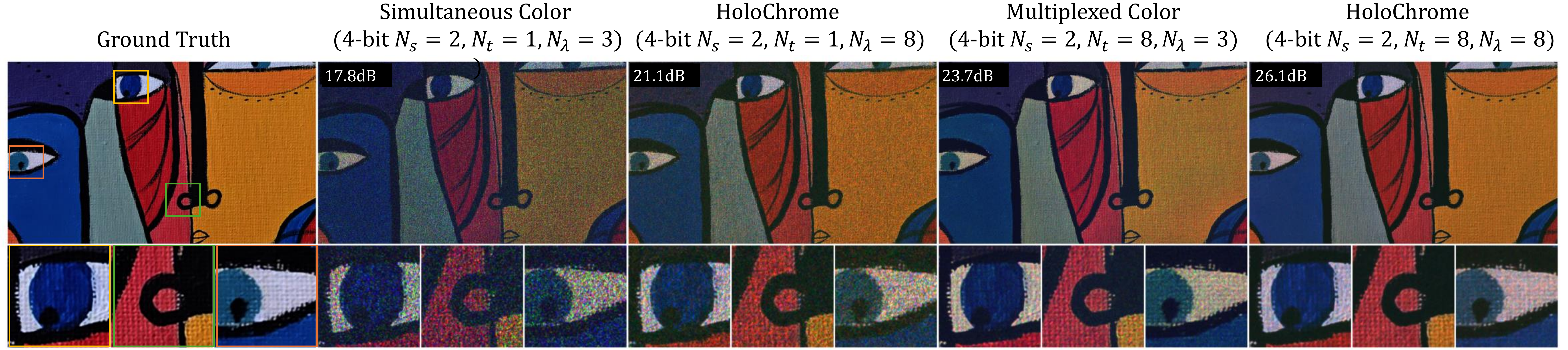}
\caption{
\textbf{HoloChrome 2D experimental results.}
    Although a simultaneous/multiplexed (3 wavelengths) random phase hologram can theoretically control speckle well for a 2D image, the experimental 2D capture has visible speckle when one zooms in.
    Our HoloChrome configuration with 8 wavelengths has noticeably reduced speckle while maintaining high-frequency features, making the fine structure of the painted canvas much more visible.
    The single frame case already works surprisingly well, which could potentially be used for field/pupil scanning approaches.
    However, capturing 8 frames provides even better results. PSNR is shown in the bottom left.
    Note that conventional approaches with 4-bit SLMs usually employ 24 ( 8 x 3) frames to display color.
    }
    
\label{fig:experimental_temporal_multiplexing_2d}
\end{figure*}

\paragraph{SLM Model}
\label{subsec:Wavelength_slm_model_calibration}

In our simulations, we use a continuous-valued phase SLM with a linear LUT. For our experiments, we use the Phase Light Modulator (PLM) device from Texas Instruments, which produces 4-bit phase modulation with a non-uniform LUT~\cite{ouyang2022evaluating}.
For each PLM, the 16-bit values are passed through a learned lookup table (LUT) which describes the mapping from digital input to phase $LUT(g; \lambda)$, as described in Sec.~\ref{sec:LUT}.
The LUT is parameterized by 16 learned coefficients (one for each possible input value).
To model the quantized behavior of the SLM, we employ a simple straight-through estimator~\cite{bengio2013estimating}.
To correctly model possible higher-order effects, we upsample the phase values by $2\times$ to avoid any problems with Nyquist sampling that might occur inside our forward model.

To model wavelength dependence, we learn LUT coefficients at a specific reference wavelength $\lambda_{ref}$.
For any other incoming field at wavelength $\lambda_{in}$, we simply scale the LUT coefficient by the ratio $\lambda_{ref} / \lambda_{in}$.
This accurately models the physics as the PLM mirrors move up/down, producing a wavelength-independent optical path difference (OPD), and a phase modulation that scales linearly with wavelength.

\paragraph{SLM Alignment} 

To address potential misalignment between the two SLMs at a sub-pixel level, we need to consider their relative positions in our model.
After the field is propagated from the first SLM, we utilize a learned warping function to transform the field into the coordinate space of the second SLM.
This warping function, derived from thin-plate spline model~\cite{duchon1977splines}, compensates for any non-radial distortion between the SLMs, allowing for precise alignment even when non-ideal relay optics are present.
The warping is executed in a differentiable manner using Kornia~\cite{riba2020kornia}, employing bilinear interpolation on both the real and imaginary components of the complex field.
We only learn one transform for all wavelength since we do not expect SLM alignment to be wavelength dependent.

\paragraph{Hyperspectral Aperture Aberration Model} To model the OPD in the aperture aberration function $\mathcal{A}(\mathbf{u}, \lambda)$ from Eqn.~\ref{eq:OPD}, we expand the 3D function into a separable basis consisting of a 2D Zernike basis in frequency coordinates $\mathbf{u}$ and a 1D polynomial in wavelength. We typically use a fairly low dimensional model for the OPD, e.g., 13 Zernike coefficients and an 8-degree polynomial in wavelength. 
We also derive a 3D adaption to allow for spatially varying aberrations, however, haven't used this in our final configuration (see supplemental material for a detailed discussion of our aberration functions).
For the amplitude component, we learn a single tensor (in our case with size of $1024^2$) at a reference wavelength $\lambda_{ref}$ (440nm for us). 
For any other incoming field at wavelength $\lambda_{in}$, we simply geometrically scale the tensor by a factor of the ratio $\lambda_{ref} / \lambda_{in}$.     

Fig.~\ref{fig:learned_holochrome_model} bottom depicts the learned amplitude and the corresponding Zernike aberrations for four selected anchor wavelengths and both relays.
We have found that our low-dimensional aperture model is very robust and can estimate aperture aberrations with extremely high fidelity. Notice the quality of the reconstructed aperture term for the second relay.
The DC-filter used for the second relay is manufactured by Thorlabs, and a small 'Thorlabs' logo is etched on the glass that is visible in our trained model parameters.
Our calibration procedure is robust enough to reconstruct fine details in the Fourier domain even though we capture images only in spatial domain.

\subsection{Calibration Procedure}
\label{sec:calibration_procedure}
To optimize the learnable parameters of our model, we gather an experimental dataset comprising SLM-image pairs and employ gradient descent optimization in PyTorch to fine-tune the unknown parameters.
We first perform a geometric alignment between both SLMs and the sensor by sequentially displaying a grid of Fresnel patterns on each SLM (and zero phase on the other) to create two grid-like calibration targets on the detector plane. Once we have established a rough alignment, we train all model parameters, including refinement of the thin-plate spline parameters used for alignment.   

Unlike many models in previous studies \cite{Choi2021NeuralDisplays, Chakravarthula2020LearnedDisplays}, our model does not incorporate any black-box neural networks and all parameters have a physical interpretation.
This approach reduces the number of learnable parameters, thereby requiring less training data, speeding up the optimization process, and minimizing the risk of overfitting.
For instance, our training dataset contains only 100 captures over the full spectrum to be calibrated, and the training process takes about 5 to 10 minutes on an Nvidia A6000.
Additionally, even though we capture training data at a single propagation distance $z$, our model generalizes well to other planes without the need for retraining.

\begin{figure*}
    \centering
    \includegraphics[width=\textwidth]{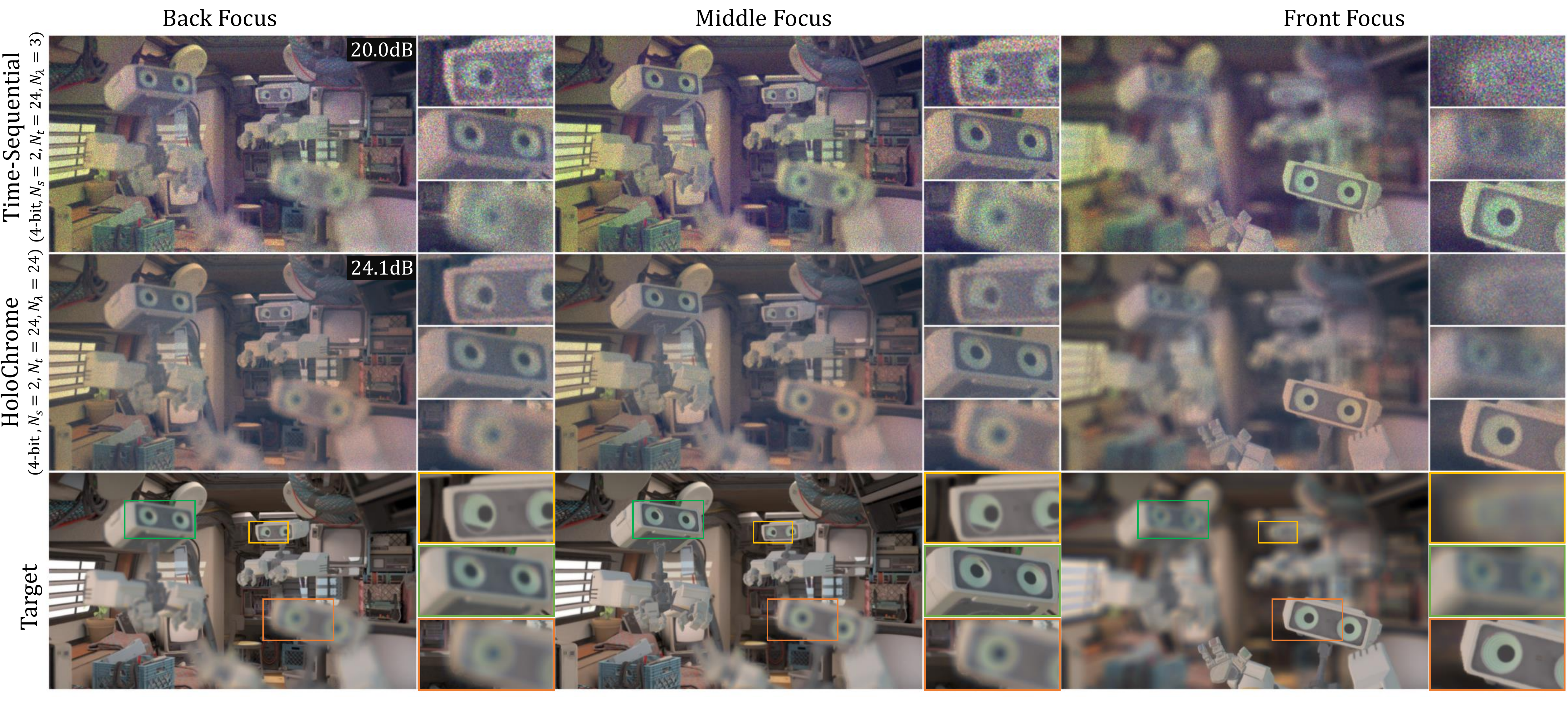}
    \caption{
       \textbf{HoloChrome experimental focal Stack results.}
       Focal stacks created by a single source hologram with random phase (top) suffer from severe speckle noise since there are insufficient degrees of freedom on the SLM to control speckle throughout a 3D volume.
       Our holochrome approach with $24$ time-multiplexed wavelengths greatly reduces speckle, enabling experimental focal stacks with natural defocus cues.
       Both focal stacks were displayed using 24 4-bit quantized frames in total.
       PSNR calculated over the full focal stack is shown on the top-left and HoloChrome is shown to increase PSNR by 4.1dB.
        }
    \label{fig:experiment_focalstack_robots}
\end{figure*}

\section{Experimental Results}
\label{sec:experiments}

\subsection{Setup Description}

We demonstrate our proposed HoloChrome setup on a benchtop experimental system, depicted in Fig.~\ref{fig:setup_detailed_sketch}.
The system employs a SuperK FIANIUM (NKT Photonics) supercontinuum laser source, which provides a broad spectrum of light from 400 nm to 2400 nm with a bandwidth around 1-2nm.
This source is coupled with a SuperK SELECT wavelength filter, allowing us to select and tune specific wavelengths across the visible spectrum with high precision.
This combination enables the polychromatic illumination central to our HoloChrome concept, providing the flexibility to explore various wavelength combinations and their effects on speckle reduction and color reproduction.

The system uses two phase-only PLMs~\cite{ouyang2022evaluating}.

A 4$f$ system with 1:1 magnification ($f_1=f_2=\SI{500}{\milli\metre}$) relays the first SLM to a distance $\Delta z = \SI{5}{\milli\metre}$ behind the second SLM.
A second 4$f$ system ($f_3=\SI{500}{\milli\metre}$, $f_4=\SI{400}{\milli\metre}$) relays the SLMs to the camera sensor. 
Irises in the Fourier planes of both 4$f$ systems filter higher orders from the SLMs.
For the second relay, we've added a DC filter to block unfiltered light stemming from the dead-time of the PLM as well as other possible sources of the DC (such as a fill factor < 1).
In future iterations, this dead-time could be mitigated by accurately synchronizing the trigger of the display and the laser pulsing.

PLM patterns are optimized using the calibrated model outlined in Sec.~\ref{sec:experiment_calibration}. 
The PLMs are initialized with uniform random patterns (4-bit), and we always jointly optimize both PLMs.

Images are captured on a monochrome camera sensor (XIMEA MC089MG-SY), mounted on a brushless translation stage (Thorlabs DDS050) to enable focal stack capture.
Focal stacks are optimized over a $10mm$ range centered at $z = 25mm$, while the actual distances at the translation stage are slightly less due to the demagnification of the second 4$f$ system.

Since the sensor is monochrome, color results are captured sequentially for each displayed wavelength and combined in post-processing using the LMS-eye response curves.
After capture, but before sending the signals through the LMS-curves, images are rectified into the PLM coordinate space using bilinear interpolation, un-modulated areas of the image are cropped out, and the relative intensities of the color channels are adjusted to ensure they have the correct optimized intensity.

A number of experimental holography papers include results with ``active'' camera-in-the-loop calibration where SLM patterns are iteratively refined by capturing the model output and then backpropagating error through a differentiable holographic display model~\cite{peng2020neural, kuo2023multisource}. This method can be very effective in improving image quality when models cannot be trained with sufficient accuracy.        
Notably, our holograms do not employ ``active'' camera-in-the-loop optimization because our learned model provides sufficiently accurate results. This not only demonstrates the robustness and accuracy of our hyperspectral calibration and modeling approach but also underscores the practicality of our system.
Removing the need for Active-CITL is critical for real-world implementation, as reliance on a feedback loop is impractical.

\subsection{2D Results}
Fig.~\ref{fig:teaser_figure} (left) shows a qualitative experimental result where a high-quality 2D image is reconstructed using HoloChrome, with significantly reduced speckle noise relative to time-sequential color.
Figure~\ref{fig:experimental_temporal_multiplexing_2d} shows a 2D experimental capture on our system at $z = \SI{25}{\milli\metre}$ comparing simultaneous/multiplexed color ($N_{\lambda}=3$) and HoloChrome ($N_{\lambda}=8$) for both single frame and time-multiplexing with $N_t=8$. While a traditional holographic display can theoretically create very high-quality 2D holograms, in practice there is still speckle noise visible in a random phase hologram. Here we demonstrate both qualitatively and quantitatively that wavelength multiplexing can significantly reduce speckle noise in experiments. 
In this example, the HoloChrome configuration improves PSNR by 3.3 dB for single frame capture and by 2.4 dB when capturing 8 frames. Note that in this 2D example, further speckle reduction is possible since we only employ 8 temporal frames while 4-bit PLMs are capable of 24.

\begin{figure*}
    \centering
    \includegraphics[width=\textwidth]{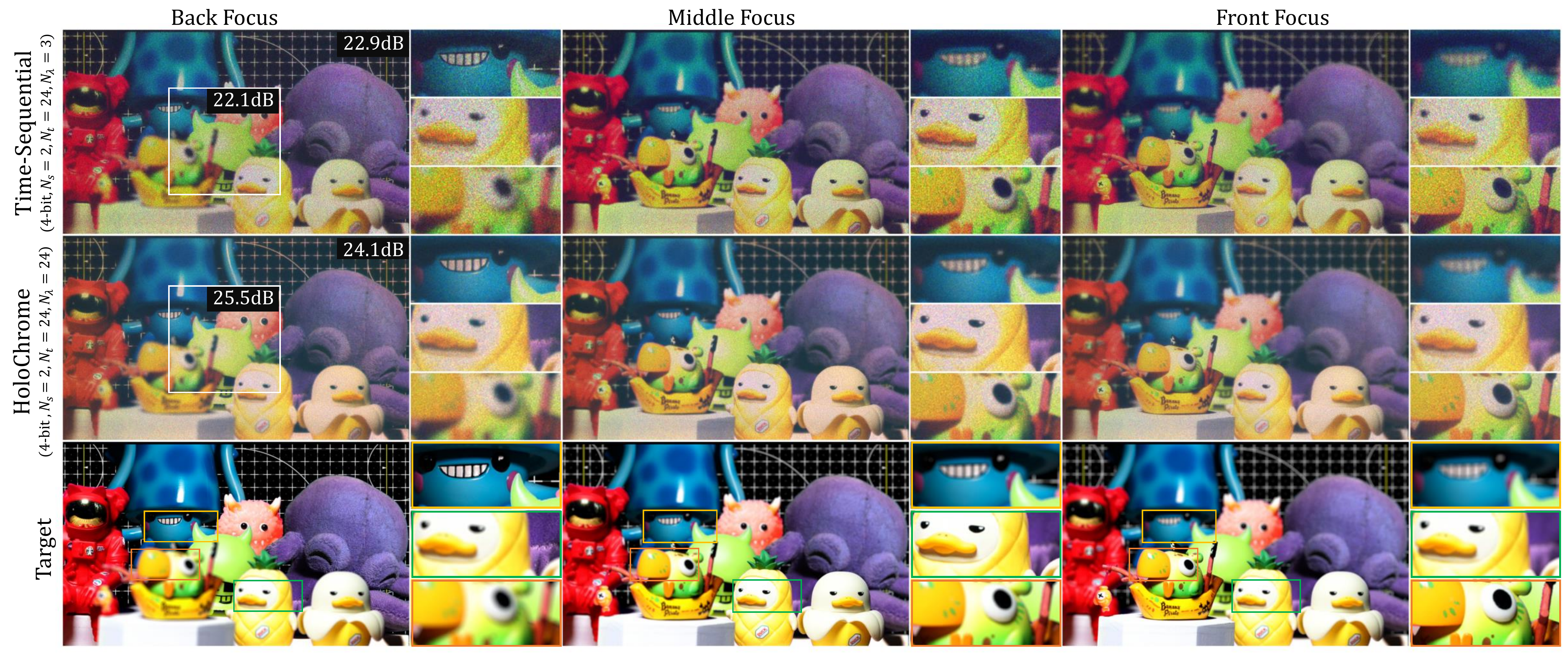}
    \caption{
       \textbf{HoloChrome experimental focal Stack results.}
       Focal stacks created by a single source hologram with random phase (top) suffer from severe speckle noise since there are insufficient degrees of freedom on the SLM to control speckle throughout a 3D volume.
       Our HoloChrome approach with $24$ time-multiplexed wavelengths greatly reduces speckle, enabling experimental focal stacks with natural defocus cues.
       Both focal stacks were displayed using 24 4-bit quantized frames in total.
       In this example, the black-level in the background is poorly reconstructed, and the PSNR boost for HoloChrome over the full frame is not significant. Despite having lower contrast, speckle is significantly reduced, and the PSNR computed over a center crop of 400x400pix that avoids the background results in a 3.4dB increase.
}
    \label{fig:experiment_focalstack_toys}
    \vspace{-.1in}
\end{figure*}

\subsection{Focal Stack Results}

Similar to Multisource Holography~\cite{kuo2023multisource}, the true benefits of HoloChrome become most apparent when displaying 3D content. Using our calibrated model, we optimize the PLM patterns while targeting a focal stack with natural blur. Fig.~\ref{fig:teaser_figure} (right) shows a qualitative example where HoloChrome reproduces a high-quality focal stack, with significantly reduced speckle noise when compared against time-sequential color.   
Figures~\ref{fig:experiment_focalstack_robots} and \ref{fig:experiment_focalstack_toys} show two examples of experimentally captured results. As expected from our simulations, the time-sequential color holograms made with $N_{\lambda}=3$ wavelengths are severely corrupted by speckle noise. In contrast, our HoloChrome approach with $N_{\lambda}=24$ significantly reduces speckle, producing high-quality images throughout the entire focal volume. This results in an improvement in PSNR of 4.1dB for the robots scene. For the toys scene, the black-level in the background is poorly reconstructed so that the PSNR over the whole frame is not improved significantly. However, while the contrast is reduced for HoloChrome, the speckle reduction is still significant, as can be observed from the insets showing closeups of the toys. In addition, a center crop of 400x400 pixels that avoids the black background exhibits a 3.4dB boost in PSNR.

%

\section{Discussion and Limitations}
This paper presents HoloChrome as an effective method for reducing speckle noise in holographic displays while maintaining image resolution and creating focal stacks with realistic blur. Our simulations and experimental results show that HoloChrome outperforms conventional methods, especially in 3D content where speckle noise is more pronounced.

A notable contribution of this work is the introduction of a hyperspectral modeling framework using polychromatic illumination and a dual-SLM architecture. This approach shows promise for future holographic displays by providing clear and high-quality images. HoloChrome works robustly for random phase holograms that produce uniform eyebox intensity (Please see supplemental material, for example, images the eyebox produced by HoloChrome). Uniform eyebox intensity is particularly beneficial for near-eye display applications where pupil position can vary significantly, reducing artifacts and enhancing the viewing experience~\cite{chakravarthula2022pupil}. However, several challenges remain to implement a practical HoloChrome near-eye system.

\paragraph{Speckle Reduction versus Contrast}
The authors of Multisource Holography~\cite{kuo2023multisource} report a tradeoff between contrast and speckle reduction that can be achieved with the method: a larger number of sources tends to increase speckle reduction at the cost of decreased contrast. We have performed similar comparisons for HoloChrome, where we study the image quality achieved as a function of the number of polychromatic wavelengths ($N_{\lambda}$) used (see Fig.~\ref{fig:ablation_study_bandwidth}), and have not observed a significant loss in contrast as $N_{\lambda}$ is increased. Furthermore, our 2D experimental results have demonstrated low black levels and high contrast (see Fig.~\ref{fig:experimental_temporal_multiplexing_2d}). However, some of our focal stack results reconstruct black levels poorly and exhibit reduced contrast (see Fig.~\ref{fig:experiment_focalstack_toys}). Further investigation is required to determine if this discrepancy is the result of modeling error in simulation, or experimental error in calibration and display (please see supplemental material for comparisons between experimental capture and predicted model output).    

\paragraph{Single versus Dual SLM}
HoloChrome achieves the best performance with a 2-SLM system. However, exploring a Single SLM HoloChrome configuration opens possibilities for simpler and more cost-effective designs. Replacing one of the SLMs with a static diffractive optical element (DOE) could still offer many benefits, although despeckling efficiency might decrease due to fewer degrees of freedom. Our simulations suggest that a single SLM setup remains effective, indicating potential for practical implementation.

\paragraph{Perceptual Studies}
\citet{markley2023simultaneous} already explored perceptual loss functions, such as those based in HSV color space.
A recent paper studied color reproduction in holography~\cite{chen2024ultrahigh}, which utilizes an active camera-in-the-loop setup to ensure precise color matching in smooth-phase holograms.
However, it is important to note that color perception in holographic near-eye displays, particularly the effect of speckle on color perception, has not been extensively studied.
Further perceptual user studies, beyond existing research on parallax or accommodation \cite{georgiou2023visual,kim2024holographic,kim2021vision}, are necessary to fully understand the implications of speckle on color perception in holographic displays.
This is particularly true, as the human vision system is not accustomed to seeing images corrupted by speckle noise, and conventional algorithms for holographic displays often assume a color-sequential 3-primary RGB representation \cite{makowski2012simple} spanning a larger color gamut compared to classical 2D color displays~\cite{zhan2020multifocal}.
Furthermore, optimizing holographic display output using the conventional 3-laser primaries and applying a mean-squared error (MSE) loss with sRGB targets may result in perceptually different colors than anticipated \cite{sharma2017digital}.
This discrepancy could be attributed to factors such as the gamma correction in sRGB images, which should be taken into account before optimizing in different color spaces~\cite{poynton2012digital}.

\paragraph{Compact and Practical Architecture}
Our HoloChrome experimental prototype is far too large to implement as a near-eye display. The dual-SLM setup and the use of super-continuum lasers or polychromatic illumination present significant hurdles in terms of miniaturization and integration into small form factors. We envision a compact architecture that integrates HoloChrome holography into near-eye displays, possibly using waveguides and transmissive amplitude modulators. However, while waveguides have been demonstrated as a feasible path towards miniaturization of near-eye holographic displays~\cite{jang2022waveguide}, further investigation is required to determine if incorporating polychromatic illumination into such a system is either feasible or practical. Developing a practical implementation for polychromatic illumination for holographic displays is a challenging task in and of itself. While implementing a polychromatic illumination module with ($>N_{\lambda}=3$) discrete laser sources is a feasible path towards implementation, it remains to be seen whether such a path is a practical approach for future commercial near-eye display architectures. 

\paragraph{Computation Speed}
Real-time computation is another significant challenge for random-phase holograms. The intensive computational requirements necessitate the development of more efficient algorithms and hardware acceleration to achieve real-time performance. Moreover, while this paper has focused on speckle reduction, future work should address color accuracy. Incorporating advanced color loss functions and perceptual models will ensure that displayed images are not only speckle-free but also accurate and vivid in color.

\section{Conclusion}
\label{sec:conclusion}
HoloChrome introduces a clear advancement of the development of high-quality, immersive holographic displays.
Our approach produces high-quality imagery for random phase holograms that produce a uniform eyebox and realistic focal cues.
To our knowledge, this is the first work to explore wavelength diversity in the field of near-eye holographic displays.
By utilizing polychromatic illumination and a dual-SLM architecture, we have shown notable improvements in speckle reduction and overall image quality, achieving results comparable to or better than existing state-of-the-art methods.
%
We believe that our work adds valuable insights into the design of future holographic displays, providing a path forward for high-quality, immersive visual experiences. However, overcoming the challenges of system complexity, miniaturization, and real-time computation will be crucial for HoloChrome to emerge as a practical solution for commercial near-eye displays.

\bibliographystyle{ACM-Reference-Format}
\bibliography{bibliography.bib}

\beginsupplement 

\title{HoloChrome: Supplemental Material}

\maketitlesup
\setcounter{page}{1}

This supplemental material includes additional derivations, implementation details, and experimental results. Please also see the supplemental videos for visualizations of the focal stack experiments and simulations.


\begin{figure}[tb]
\centering
\includegraphics[width=1.0\linewidth]{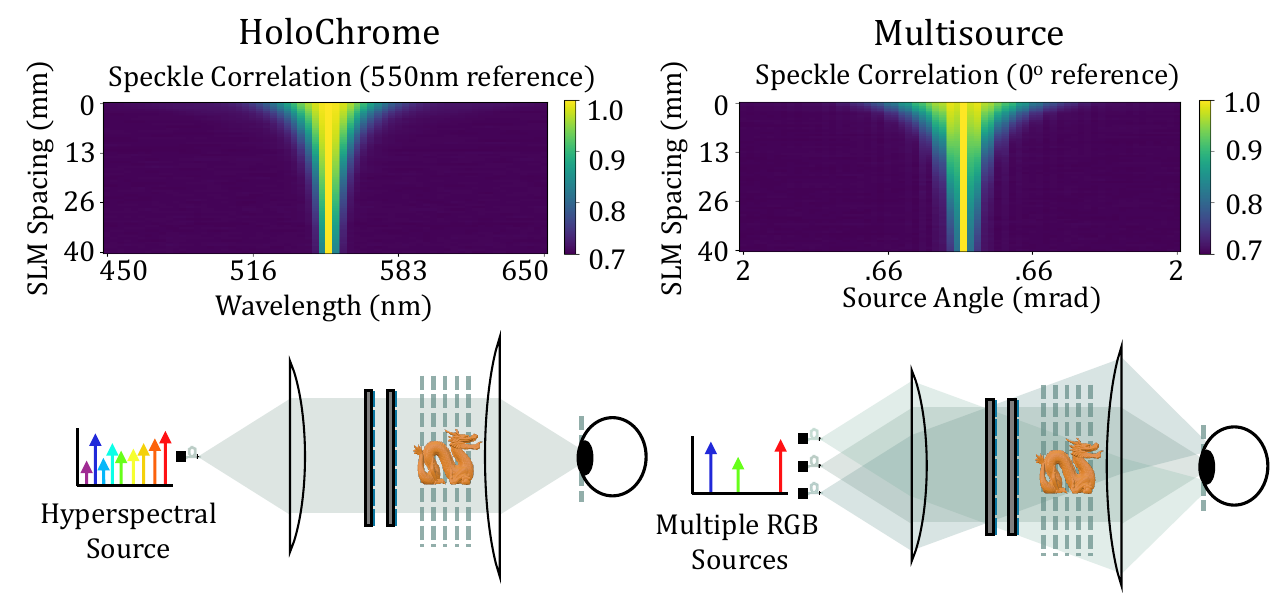}
\caption{
\textbf{Speckle correlation for both HoloChrome and Multisource.} 
HoloChrome and Multisource holography use orthogonal methods to create independent speckle patterns but exhibit very similar behavior.
In this simulation, we use a two-SLM configuration for both HoloChrome and Multisource ($N_s=2$) and choose a random, static pattern on both SLMs.
We first change the wavelength and the incoming field angle for both setups respectively, and also change the spacing between the two SLMS to efficiently break the memory effect as outlined in~\cite{kuo2023multisource}. The speckle correlations are near identical for the chosen parameters, indicating that the dual-SLM architecture behaves similar for both wavelength and angle multiplexing of source illumination. 
}
\label{fig:simulation_correlation_plot}
\Description{Simulation Correlation Plot}
\end{figure}

\section{Wavelength Memory Effect}
\label{sec:Wavelength_Memory_Effect}

The \textbf{wavelength memory effect}~\cite{willomtizer2021} describes how speckle patterns produced by coherent light interacting with an optical system change as the wavelength varies. Understanding this effect is crucial for determining the minimal wavelength difference required to generate uncorrelated speckle patterns at different wavelengths, which is essential in applications like HoloChrome.

Figure~\ref{fig:simulation_correlation_plot} illustrates the direct analogy between the decorrelation effects between angular and wavelength multiplexing.
The cross-correlation function between a speckle pattern created at the center wavelength (or center source angle) and a variation from it is plotted.
In Multisource Holography, we change the illumination angle; in HoloChrome, we change the wavelength.
Yet, both methods clearly exhibit similar behavior, demonstrating that wavelength multiplexing effectively achieves speckle decorrelation in an orthogonal way compared to Multisource.

\section{HoloChrome Simulation Details}
\subsection{Upsampling SLM to avoid aliasing}
We simulate at double the SLM resolution in both dimensions (i.e., 4 simulation pixels per SLM pixel) to prevent aliasing. 
Aliasing arises because the spatial frequency of the signal doubles when converting to intensity, and without upsampling, high-frequency signals may be misrepresented. 
This phenomenon, often overlooked in holographic displays, is significant, particularly for multi-display holograms like ours. 
For more details, we refer to the argumentation in \textit{Multisource Holography}~\cite{kuo2023multisource}, as we have followed the procedure outlined in this paper.

\section{Hyperspectral Image Processing and Normalization}

In this section, we describe the procedure used for processing hyperspectral images, including rectification, normalization, and per-channel intensity correction.
\subsection{Image Loading and Preprocessing}
The hyperspectral image cube is loaded for both the model output and the captured images.
We then use the expected model output to "normalize" the captured monochromatic images to have the correct energy per channel.
Rectification of the captured images is performed using our inverse "deformation method" with the thin-plate-spline method, which warps the captured image into the SLM frame.
Subsequently, a crop operation is also performed to crop out any boundary artifacts of the SLM.

\subsection{Per-Channel Normalization}

We used a supercontinuum laser where the output laser power is a highly non-linear function of wavelength. In~\ref{sec:laser_power_calibration} we detail the procedure used to to calibrate the laser vs. wavelength. Unfortunately, the laser we used offered limited precision in wavelength dependent power tuning. As a result, we were not able to achieve the desired polychromatic source profiles directly in hardware.

Instead, we decided to simply "normalize" the output power of the measured image at a specific wavelength using the "ideal" image for that specific wavelength as produced by the polychromatic model output.
We call this a per-wavelength normalization. This ensures that the mean intensity of the rectified images matches that of the target polychromatic field, for each source wavelength used.
The final step involves computing the RGB representation of the hyperspectral data by mapping the hyperspectral cube through the physical LMS-response curves of the eyes.

\section{Additional Calibration Details}

\subsection{Alignment between both SLMs}
To calibrate the relative alignment between our 2-SLMs, we follow the warping procedure outlined in the supplementary material of \textit{Multisource Holography}~\cite{kuo2023multisource}.
To map the complex field between SLM1 and SLM2, we apply a thin-plate spline (TPS) model.
This warping is calibrated by optimizing the SLM patterns to create asymmetric dot patterns, which account for image flips introduced by the optical system.
The TPS transformation between SLMs is further fine-tuned using gradient descent alongside the other model parameters during our calibration procedure.

\subsection{Calibration Patterns and Fine-Tuning}

To calibrate the system, we generate SLM patterns by starting with a uniform random phase and applying a Gaussian kernel with variable standard deviations to introduce different levels of blur.
These SLM patterns are normalized to cover the full phase range of our SLMs(0-15).
The blurring process creates a variety of feature sizes in the images, providing a diverse dataset necessary for calibrating different parameters of the model.
Low-frequency images, generated with a larger Gaussian kernel, are primarily used to calibrate the TPS warping.
Higher frequency images are utilized for calibrating lens aberrations and the SLM LUT.

\section{Laser Power Calibration}
\label{sec:laser_power_calibration}
\begin{figure}[tb]
\centering
\includegraphics[width=1.0\linewidth]{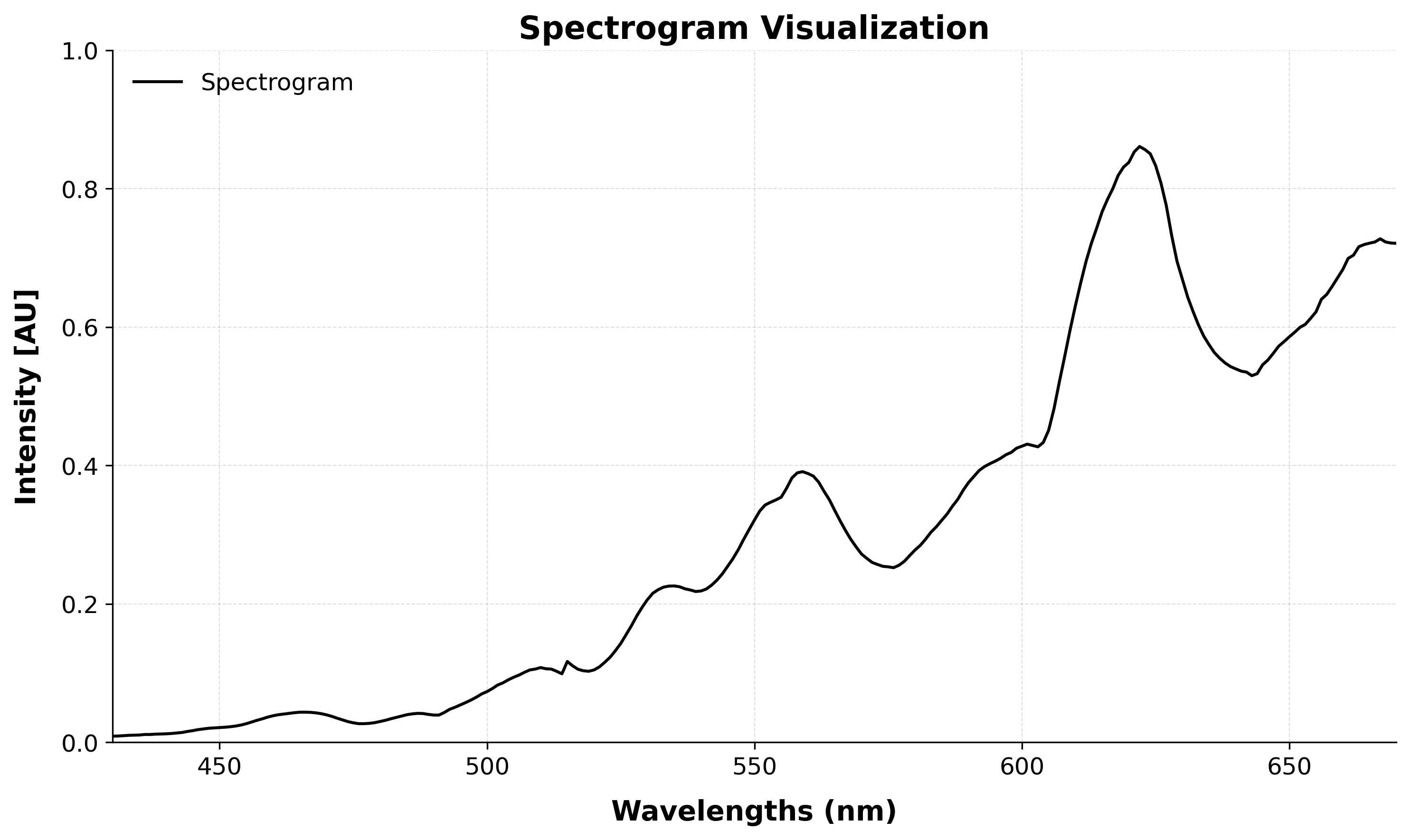}
\caption{Calibrated power spectrum of the NKT SuperK laser, showing the large intensity disparity between red and blue wavelengths.}
\label{fig:spectrogram}
\end{figure}

The power spectrum of the NKT SuperK supercontinuum laser is highly non-uniform across its wavelength range. This non-uniformity becomes even more pronounced when using the K-Select filter to isolate specific wavelengths. Figure \ref{fig:spectrogram} shows the calibrated spectrum captured on the camera after passing through the entire optical system, with both phase light modulators (PLMs) displaying speckle patterns.

\begin{figure}[tb]
\centering
\includegraphics[width=1.0\linewidth]{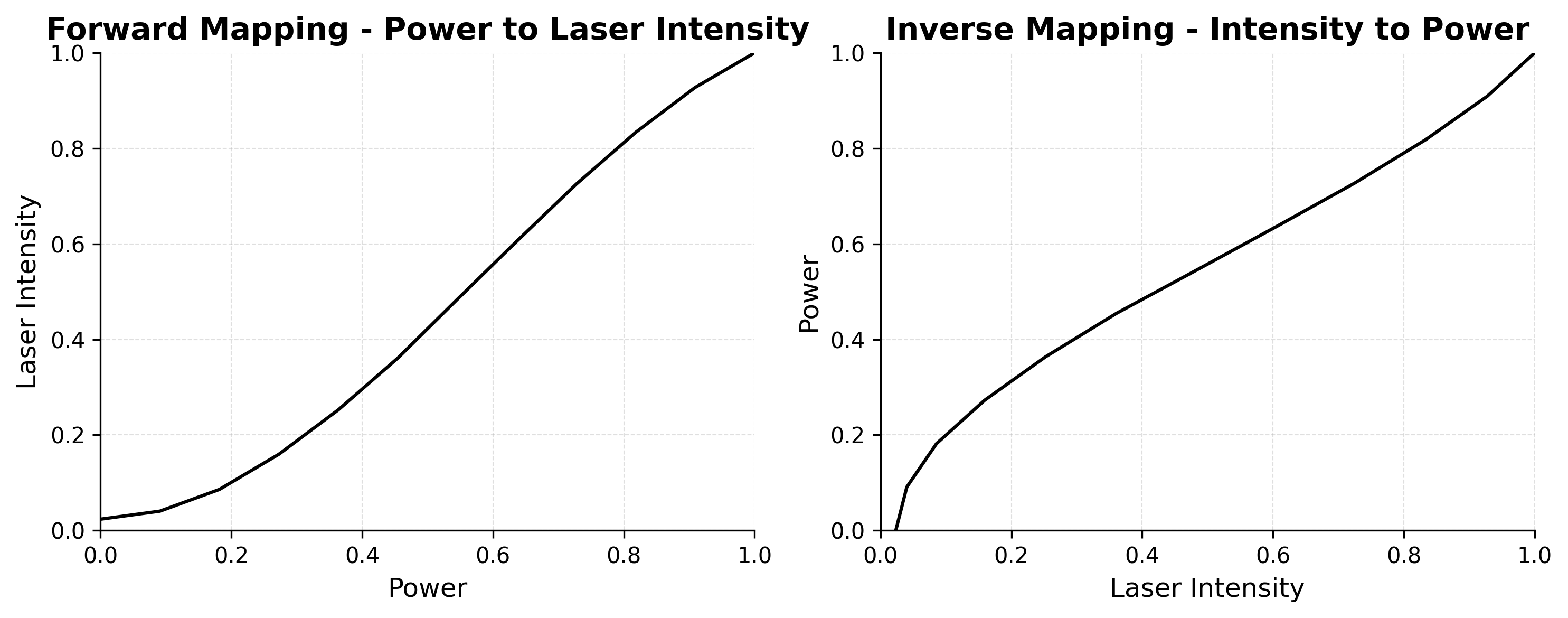}
\caption{Laser power response mapping, illustrating the nonlinear behavior at low power settings.}
\label{fig:laser_lut}
\end{figure}

As observed, the average intensity in the red wavelength range exceeds that in the blue range by a factor of 20 to 40. This presents a significant challenge for light equalization, as the dynamic range of the NKT laser is constrained. Additionally, the power response of the K-Select filter is nonlinear, especially at low power settings, see Fig~\ref{fig:laser_lut}. Specifically, when operating at lower power values, the nonlinear behavior becomes more pronounced, complicating efforts to achieve uniform illumination.

The large intensity disparity between the red and blue wavelengths, combined with the limited dynamic range of the laser, made it difficult to display wavelengths with uniform spectral response. This resulted in frequent issues with underexposure and overexposure, particularly during speckle pattern experiments used for calibration of the optical setup. Ideally, well-exposed speckle patterns were required for calibration, but the constraints imposed by the available power necessitated adjustments in the camera exposure time instead.

To address this, we optimized the system by adjusting the exposure time of the camera in relation to the frame rate of the PLM hardware. Since the PLMs operate at 1440 Hz (24 frames per 60 Hz cycle) with approximately 20\% deadtime, it was crucial to synchronize data capture at multiples of 16.6 Hz to avoid temporal aliasing issues. Based on the unpredictable dynamic range of the laser, we precomputed pairs of exposure times (multiples of 16 Hz) and corresponding laser power settings, ensuring that each image captured at each wavelength exhibited consistent image intensity statistics.

\section{Experimental Setup}
\begin{figure*}
\includegraphics[width=1.0\textwidth]{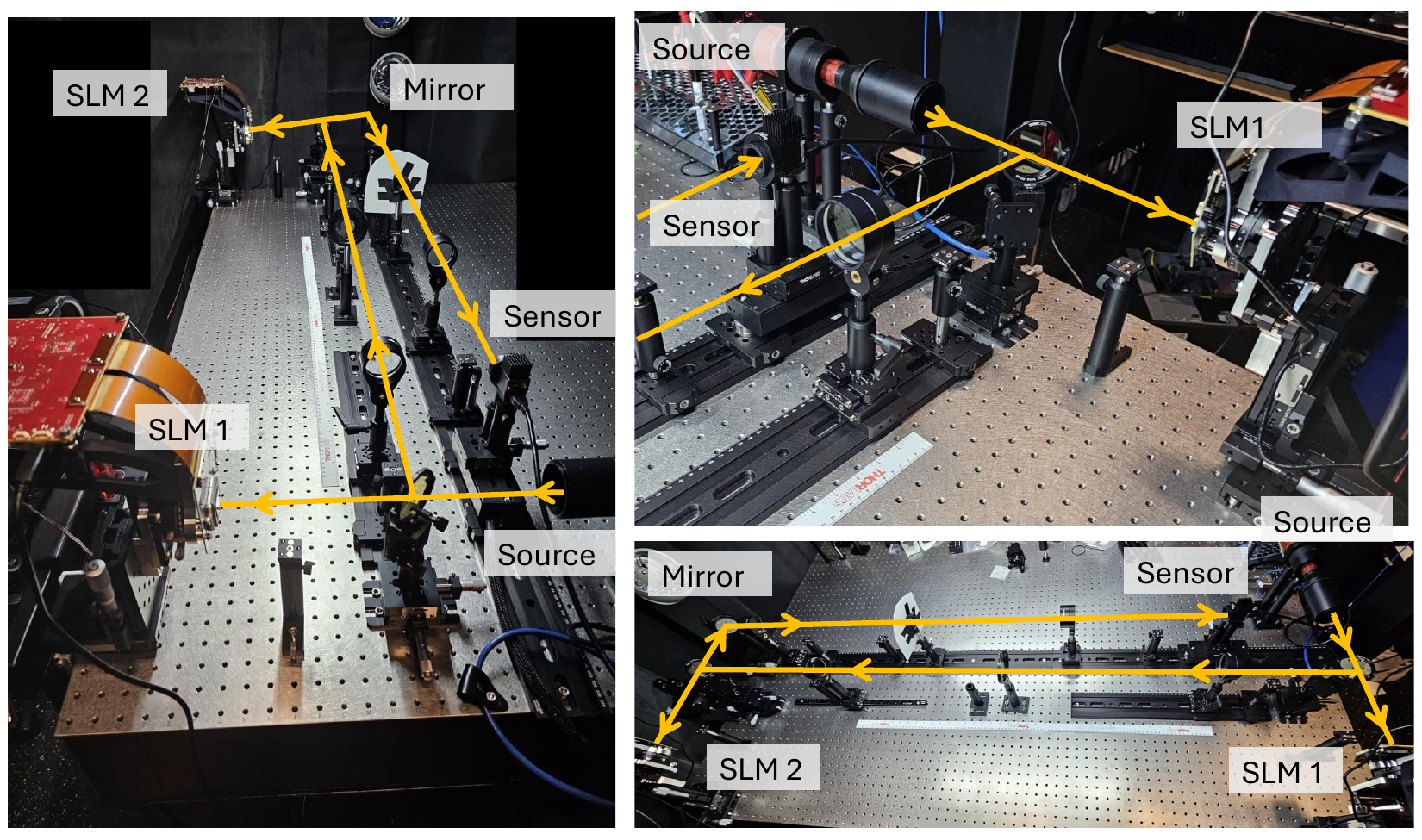}
\caption{
\textbf{HoloChrome 2D experimental setup:} 
This figure illustrates the optical setup used in the HoloChrome 2D experiment. The setup includes two spatial light modulators (SLM1 and SLM2), arranged in a $4f$ configuration with a relay system. The light source, although not visible in this image due to its large size, is a SuperK supercontinuum laser that illuminates the system with a tunable spectrum of wavelengths. The mirrors direct the light between the components, while the sensors capture the images after modulation through the SLMs. The system operates to time-multiplex different wavelengths for holographic image capture.
}
\label{fig:experimentalsetup}
\end{figure*}

The experimental setup for the HoloChrome system is shown in Fig.~\ref{fig:experimentalsetup}. The system utilizes two spatial light modulators (TI-PLM), SLM1 and SLM2, arranged in a $4f$ relay configuration. The laser source, a SuperK supercontinuum laser, provides a broad spectrum of wavelengths but is not shown in the figure due to its large size.
The light is guided through a beamsplitter to the SLMs, which are used to modulate the phase of the incoming light.
To reduce spatially varying aberrations, we use large focal lengths (500mm) for both relays. The last lens is 400mm leading to a slight demagnification of the system.
Note that, the propagation distance (or travel distance on the translation stage where the camera is mounted) has to be scaled accordingly with this magnification factor.
Images are captured on a monochrome sensor after the light has passed through the system, with the sensor positioned to record the multiplexed holographic data for each wavelength.
The setup allows for flexible tuning of wavelengths to capture multispectral holographic images.

\section{Modeling Spatially Varying Aberrations in Hyperspectral Holography}
\label{supp:subsec:modeling_spatial_varying_aberrations}
Accurately modeling and compensating for spatially varying aberrations is essential to achieve high-quality reconstructions across the full spectrum of wavelengths used in hyperspectral illumination.
Traditional aberration models often assume shift-invariance across the optical field, which fails to capture the intricate variations in aberrations.
To address this limitation, we propose a sophisticated model that captures the dependency of aberrations on both spatial coordinates and wavelength.

Note that we trained our model using this spatially varying aberration model.
However, since we chose to use 500mm focal lengths lenses for our relay, the amount of spatially varying blur is very small.
Hence, we didn't use the SVA approach that is outlined in the following for our experimental setup.
However, it might be required for more compact setups. A similar approach employing the cropped-SVA has already been shown useful in \cite{kuo2023multisource}.

\subsection{Mathematical Formulation}
The goal is to model a function that maps from a 5D space (spatial frequencies $f_u, f_v$, spatial coordinates $u, v$, and wavelength $\lambda$) to a scalar representing the Optical Path Difference (OPD) as
\begin{equation}
OPD(f_u, f_v | u, v, \lambda) 
\label{eq:zernike_coefficients_opd}
\end{equation}

Zernike polynomials are chosen to model the aberrations due to their orthogonality and relevance in optical systems.
Since computing the aberration map from a set of zernike coefficients, we can reduce learning a function that maps a point form $R^3$ ($u, v, \lambda$) to an N-dimensional space representing the N Zernike coefficients.

The aberration model is defined as:
\begin{equation}
Z_n(u, v, \lambda) = \sum_{i=0}^{M} \sum_{j=0}^{P} c_{nij} \cdot \psi_i(u, v) \cdot \phi_j(\lambda)
\label{eq:zernike_coefficients_abberation_model}
\end{equation}
where $Z_n(u, v, \lambda)$ represents the n-th Zernike coefficient at spatial coordinates $u, v$ and wavelength $\lambda$. $\psi_i(u, v)$ are basis functions defined over the spatial domain, typically chosen as polynomials or another suitable set of functions that capture the spatial complexity. $\phi_j(\lambda)$ are polynomial terms that model the wavelength dependency of the Zernike coefficients, and $c_{nij}$ are the coefficients to be learned from the data.

From the N Zernike coefficients, the OPD at frequency coordinates $f_u, f_v$ in the angular domain can be computed as:

\begin{equation}
\text{OPD}(f_u, f_v, u, v, \lambda) = \sum_{n=0}^{N} Z_n(u, v, \lambda) \cdot \zeta_n(f_u, f_v)
\label{eq:opd}
\end{equation}
where $\zeta_n(f_u, f_v)$ represents the n-th Zernike polynomial evaluated at frequency coordinates $f_u, f_v$.

\textit{Polynomial Approximation:}
To efficiently approximate the function that maps from the 3D space ($u, v, \lambda$) to the N Zernike coefficients, a polynomial approach is employed. The spatial and spectral dependencies of the Zernike coefficients are modeled using polynomial basis functions, as shown in Eq.~\ref{eq:zernike_coefficients_abberation_model}.
The coefficients $c_{nij}$ are learned from the data, allowing the model to capture the intricate variations in aberrations across the field of view and wavelength range. This polynomial approximation enables fast computation of the Zernike coefficients and, consequently, the OPD at any given spatial position, wavelength, and frequency coordinates.

The proposed aberration model, based on Zernike polynomials and polynomial approximation, provides a computationally efficient way to accurately represent the spatially and spectrally varying aberrations in the HoloChrome system. By incorporating this model into the hologram optimization process, the system can effectively compensate for these aberrations, leading to high-quality, aberration-corrected holographic reconstructions across the wide range of wavelengths used in the hyperspectral illumination.

\section{Additional Results (Simulation)}

\begin{figure*}[htb]
\centering
\includegraphics[width=1.0\textwidth]{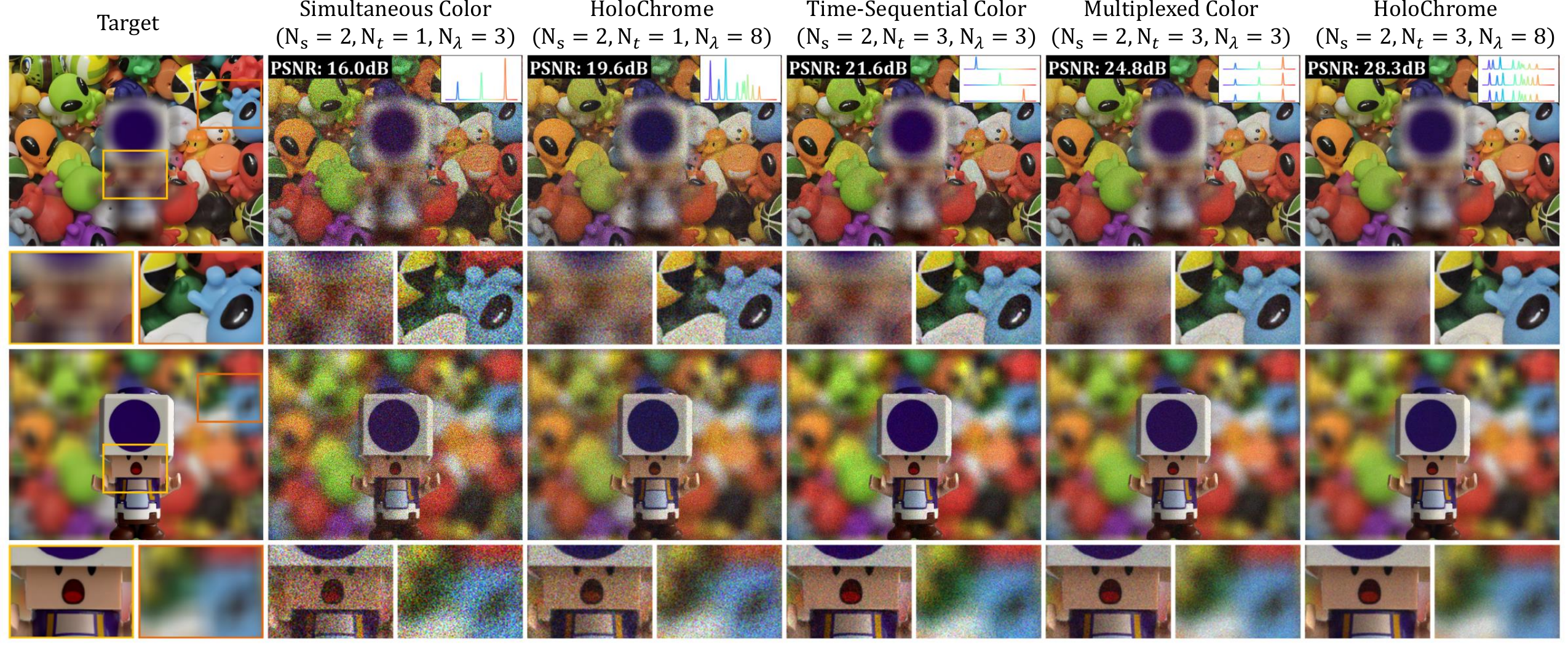}
\caption{
\textbf{Comparison of HoloChrome with conventional holography methods (simulation).}
HoloChrome shows significant performance gains in speckle reduction compared to conventional methods.
The figure compares the results of HoloChrome against three common holographic methods:
Simultaneous Color, Time-Sequential Color, and Multiplexed Color.
The PSNR values indicate that HoloChrome, both in single-frame and three-frame configurations, achieves higher image quality, with the three-frame HoloChrome providing over a 6dB improvement compared to conventional time-sequential color.
Despeckled results are evident in the insets. Illumination spectra are shown in the top right of each column.
}
\label{fig:supp:motivation_1_slm}
\end{figure*}

In Fig.~\ref{fig:supp:motivation_1_slm} we show an additional result with a different scene for our Motivation figure that we used in our main body of the paper.
The findings/results of this scene are in line with what we report in the main paper.

\subsection{Additional 3D Result (Experiment)}
\begin{figure*}
    \centering
    \includegraphics[width=\textwidth]{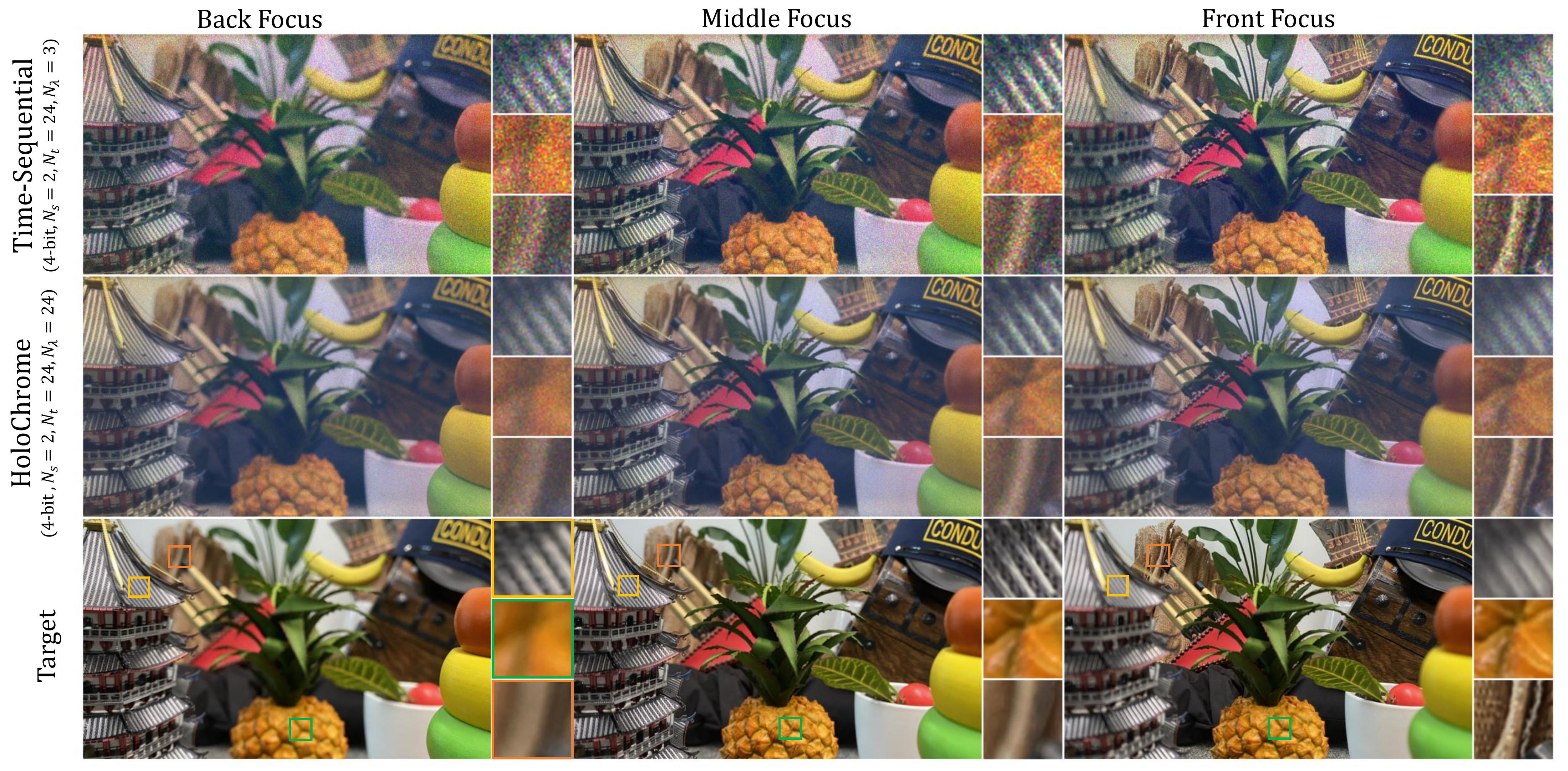}
    \caption{
       \textbf{HoloChrome experimental focal Stack results.}
       Focal stacks created by a single source hologram with random phase (top) suffer from severe speckle noise since there are insufficient degrees of freedom on the SLM to control speckle throughout a 3D volume.
       Our holochrome approach with $8$ time-multiplexed wavelengths greatly reduces speckle, enabling experimental focal stacks with natural defocus cues.
       Both focal stacks were displayed using 24 4-bit quantized frames in total.
       In this example the black-level in the background is poorly reconstructed, and the PSNR reported is computed over a crop of 400x400pix at the center of the full focal-stack.
}
    \label{fig:experiment_focalstack_flo2}
    \vspace{-.1in}
\end{figure*}

\begin{figure*}
    \centering
    \includegraphics[width=\textwidth]{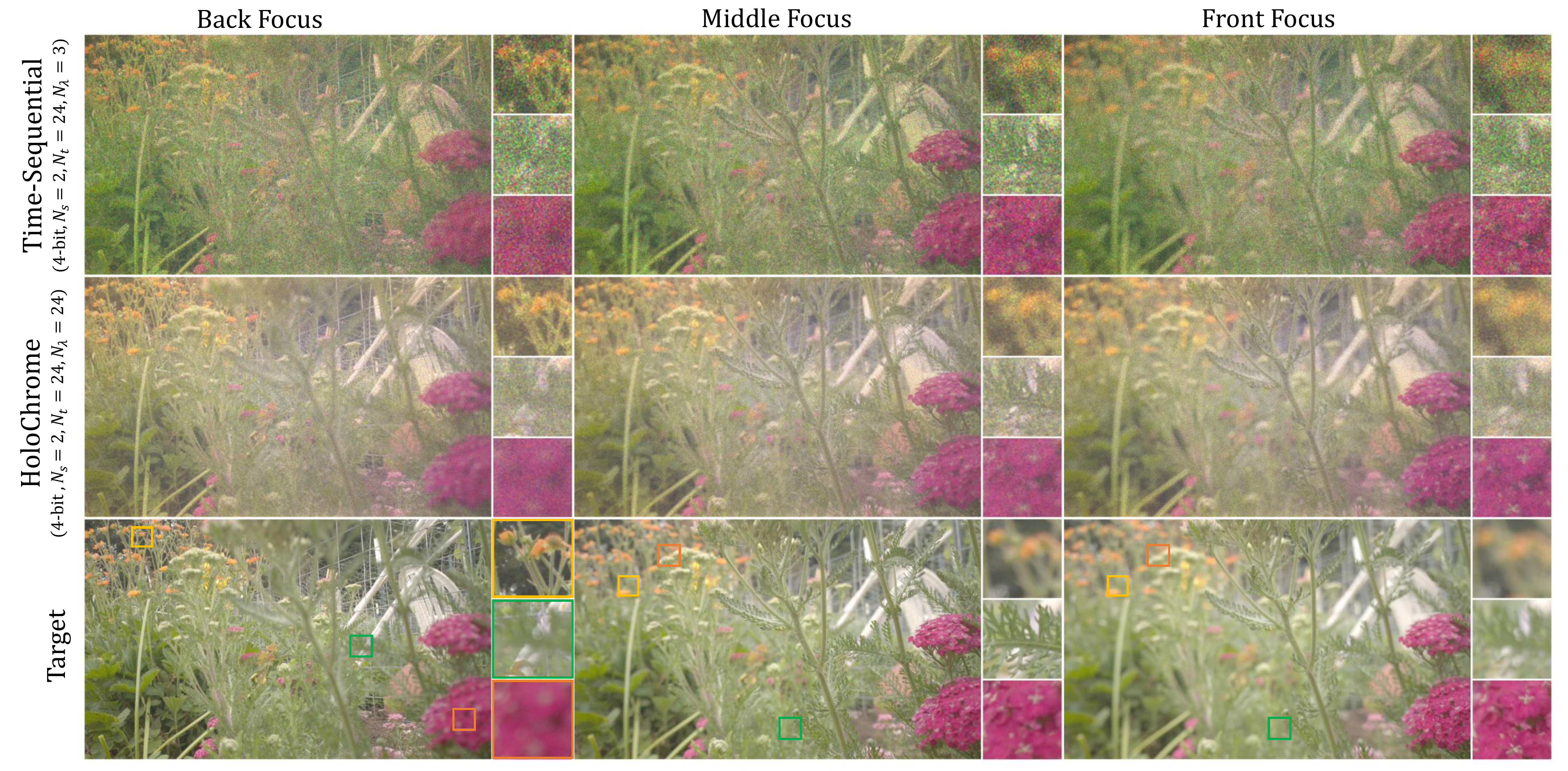}
    \caption{
       \textbf{HoloChrome experimental focal Stack results.}
       Focal stacks created by a single source hologram with random phase (top) suffer from severe speckle noise since there are insufficient degrees of freedom on the SLM to control speckle throughout a 3D volume.
}
    \label{fig:experiment_focalstack_merrymoor1}
    \vspace{-.1in}
\end{figure*}

\begin{figure*}
    \centering
    \includegraphics[width=\textwidth]{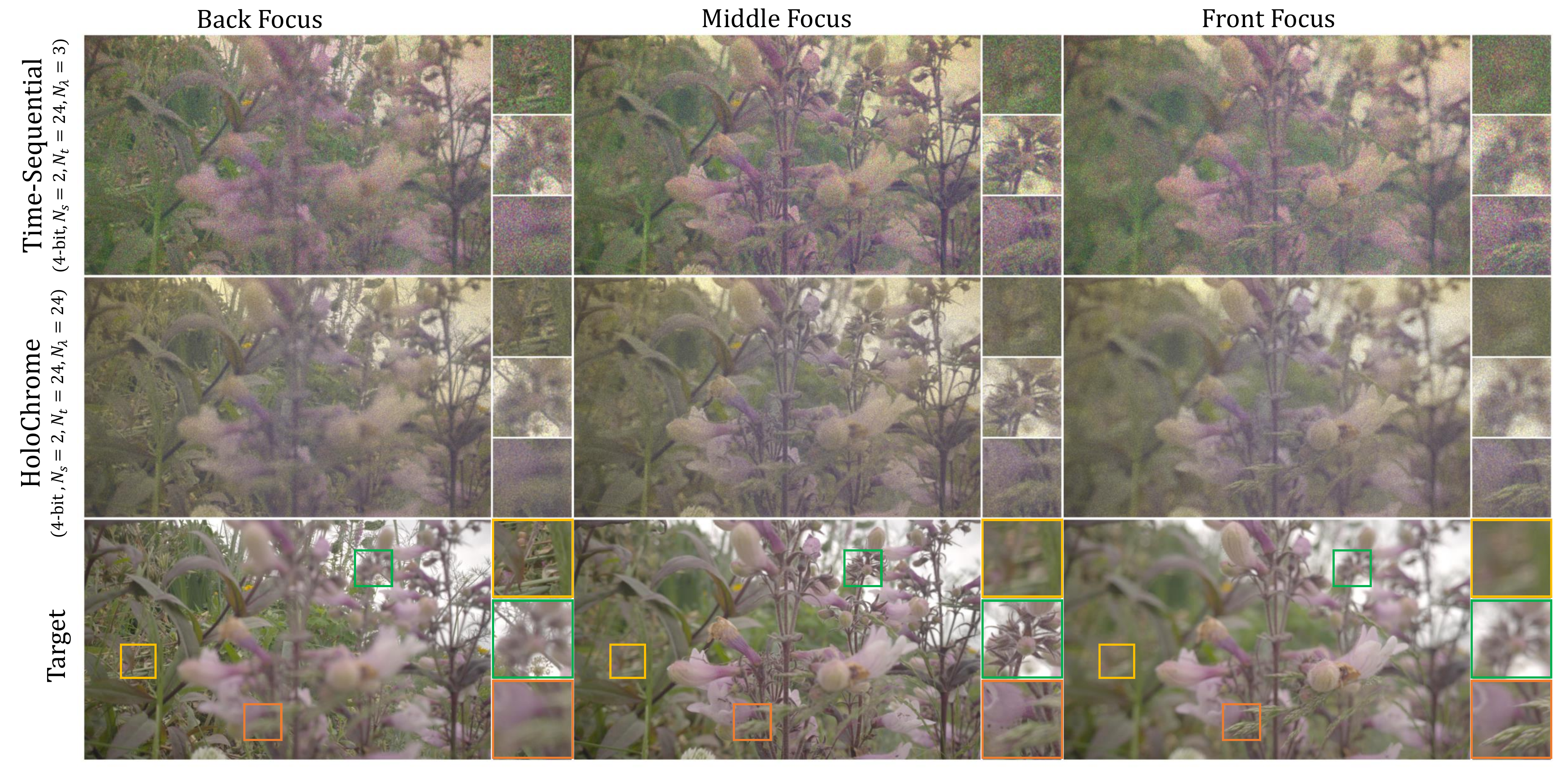}
    \caption{
       \textbf{HoloChrome experimental focal Stack results.}
       Focal stacks created by a single source hologram with random phase (top) suffer from severe speckle noise since there are insufficient degrees of freedom on the SLM to control speckle throughout a 3D volume.
}
    \label{fig:experiment_focalstack_iphone2}
    \vspace{-.1in}
\end{figure*}

\begin{figure*}
    \centering
    \includegraphics[width=\textwidth]{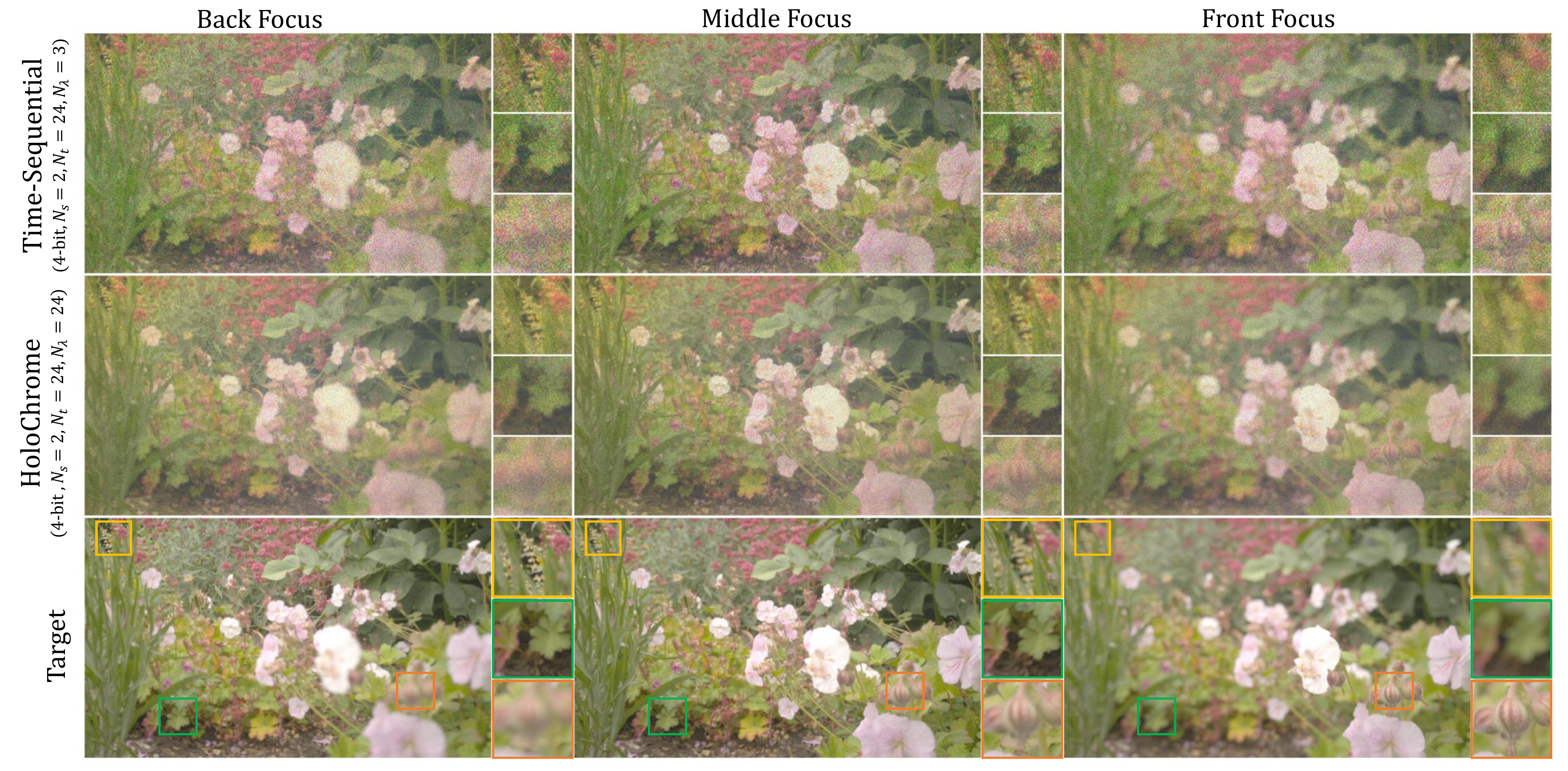}
    \caption{
       \textbf{HoloChrome experimental focal Stack results.}
       Focal stacks created by a single source hologram with random phase (top) suffer from severe speckle noise since there are insufficient degrees of freedom on the SLM to control speckle throughout a 3D volume.
       Our holochrome approach with $8$ time-multiplexed wavelengths greatly reduces speckle, enabling experimental focal stacks with natural defocus cues.
       Both focal stacks were displayed using 24 4-bit quantized frames in total.
       In this example the black-level in the background is poorly reconstructed, and the PSNR reported is computed over a crop of 400x400pix at the center of the full focal-stack.
}
    \label{fig:experiment_focalstack_iphone14}
    \vspace{-.1in}
\end{figure*}

In addition to the 2D experimental results, we provide a series of 3D focal stack results captured using our HoloChrome system. Figures \ref{fig:experiment_focalstack_flo2}, \ref{fig:experiment_focalstack_iphone14}, \ref{fig:experiment_focalstack_iphone2}, and \ref{fig:experiment_focalstack_merrymoor1} demonstrate the effectiveness of our approach in reducing speckle noise across different focal planes while preserving natural defocus cues. 

These figures show focal stacks captured with a smartphone, showcasing similar blur parameters and correct image statistics as desired for these experiments. Each focal stack was generated by time-multiplexing 8 wavelengths, which significantly reduces speckle noise compared to a single source hologram with random phase. The images are displayed using 24 4-bit quantized frames.

\subsection{Additional 2D Results for HoloChrome}

\begin{figure*}[tb]
\centering
\includegraphics[width=1.0\textwidth]{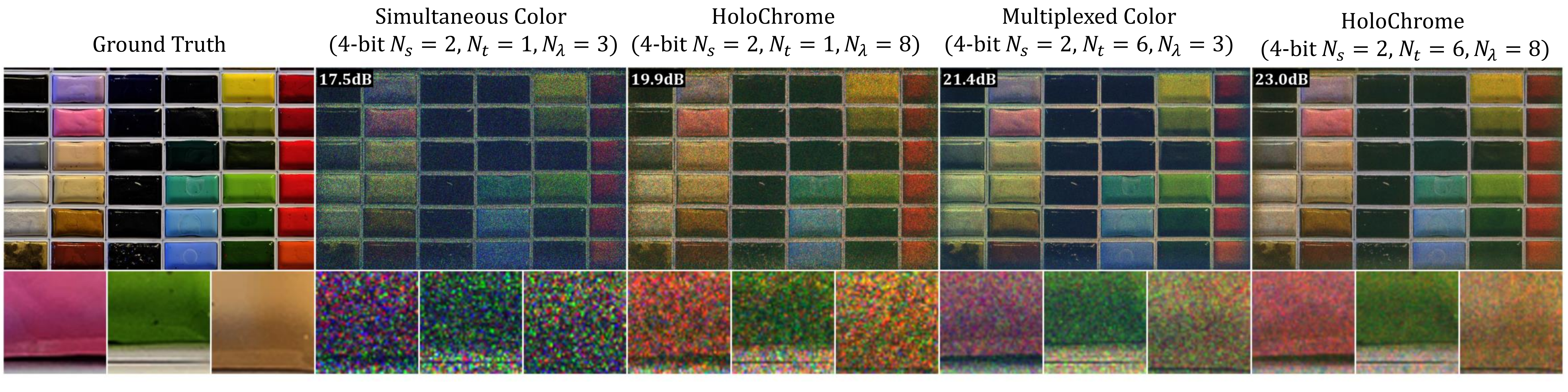}
\caption{
\textbf{HoloChrome 2D experimental results.}
    Although a simultaneous/multiplexed (3 wavelengths) random phase hologram can theoretically control speckle well for a 2D image, the experimental 2D capture has visible speckle when one zooms in.
    Our HoloChrome configuration with 8 wavelengths has noticeably reduced speckle while maintaining high-frequency features, making the fine structure of the paint canvas much more visible.
    The single frame case already works surprisingly well, which could potentially be used for field/pupil scanning approaches.
    However, capturing 8 frames provides even better results. PSNR is shown in the bottom left.
    Note that conventional approaches with 4-bit PLMs usually employ 24 ( 8 x 3) frames to display color.
    }
    
\label{fig:experimental_temporal_multiplexing_2d_colorchart}
\end{figure*}

\begin{figure*}[tb]
\centering
\includegraphics[width=1.0\textwidth]{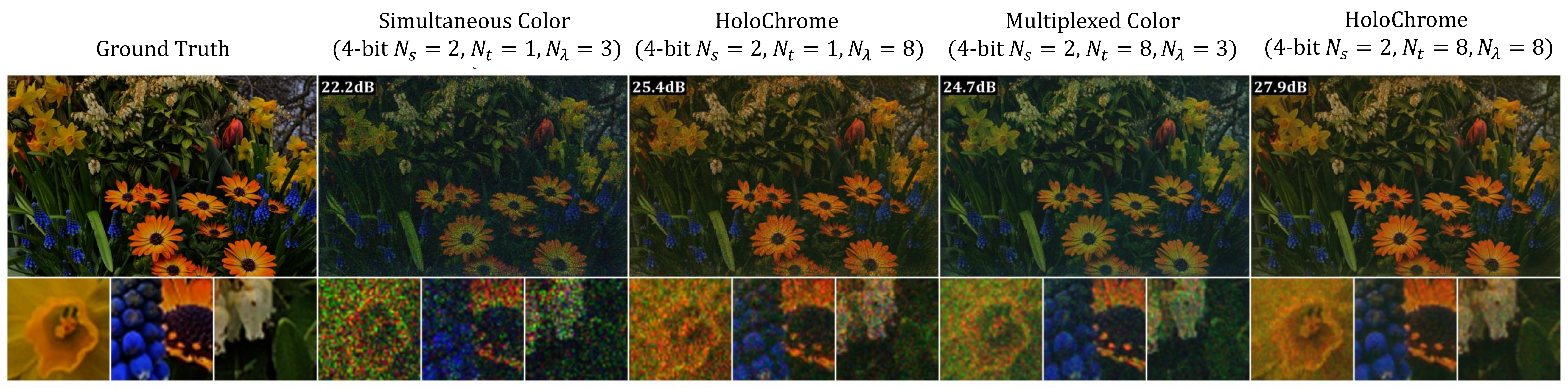}
\caption{
\textbf{HoloChrome 2D experimental results.}
    Although a simultaneous/multiplexed (3 wavelengths) random phase hologram can theoretically control speckle well for a 2D image, the experimental 2D capture has visible speckle when one zooms in.
    Our HoloChrome configuration with 8 wavelengths has noticeably reduced speckle while maintaining high-frequency features, making the fine structure of the paint canvas much more visible.
    The single frame case already works surprisingly well, which could potentially be used for field/pupil scanning approaches.
    However, capturing 8 frames provides even better results. PSNR is shown in the bottom left.
    Note that conventional approaches with 4-bit PLMs usually employ 24 ( 8 x 3) frames to display color.
    }
    
\label{fig:experimental_temporal_multiplexing_2d_flowers}
\end{figure*}

\begin{figure*}[tb]
\centering
\includegraphics[width=1.0\textwidth]{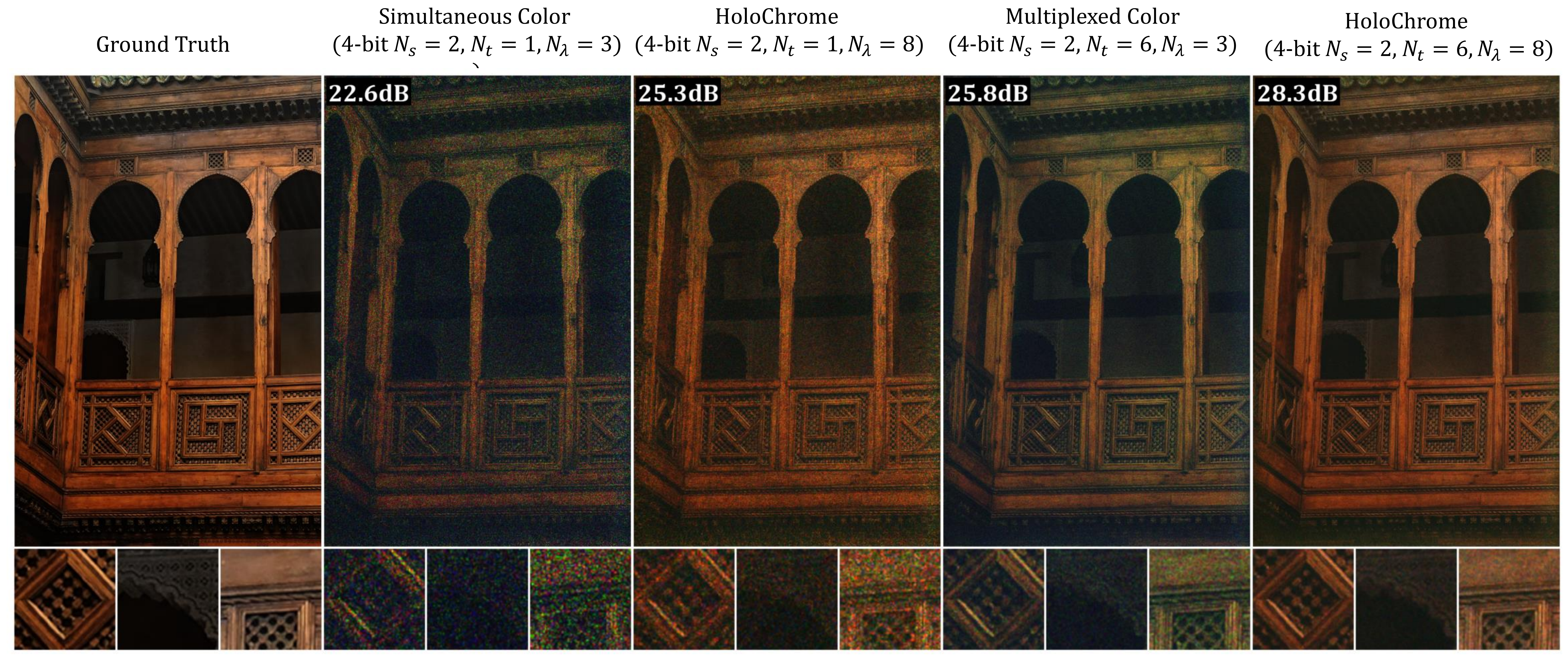}
\caption{
\textbf{HoloChrome 2D experimental results.}
    Although a simultaneous/multiplexed (3 wavelengths) random phase hologram can theoretically control speckle well for a 2D image, the experimental 2D capture has visible speckle when one zooms in.
    Our HoloChrome configuration with 8 wavelengths has noticeably reduced speckle while maintaining high-frequency features, making the fine structure of the paint canvas much more visible.
    The single frame case already works surprisingly well, which could potentially be used for field/pupil scanning approaches.
    However, capturing 8 frames provides even better results. PSNR is shown in the bottom left.
    Note that conventional approaches with 4-bit PLMs usually employ 24 ( 8 x 3) frames to display color.
    }
    
\label{fig:experimental_temporal_multiplexing_2d_morroco}
\end{figure*}

To further demonstrate the capabilities of our HoloChrome 2D system, we provide additional experimental results in Figures \ref{fig:experimental_temporal_multiplexing_2d_colorchart}, \ref{fig:experimental_temporal_multiplexing_2d_flowers}, and \ref{fig:experimental_temporal_multiplexing_2d_morroco}. These figures present different 2D captures using our HoloChrome setup with 8 wavelengths, illustrating the effectiveness of our approach in reducing speckle while preserving high-frequency details.

Figure \ref{fig:experimental_temporal_multiplexing_2d_colorchart} displays a color chart, where the speckle reduction achieved by our method enhances the visibility of fine details. Similarly, Figure \ref{fig:experimental_temporal_multiplexing_2d_flowers} shows a detailed flower arrangement, emphasizing how our method improves the structural clarity of the scene. Finally, Figure \ref{fig:experimental_temporal_multiplexing_2d_morroco} presents a complex textured surface, further demonstrating the benefits of using 8 wavelengths in our HoloChrome setup.

In all cases, the single frame captures already show promising results, potentially useful for applications such as field or pupil scanning. However, our 8-frame configuration consistently provides superior image quality, with enhanced speckle control and better preservation of high-frequency features. The PSNR values, shown in the bottom left of each figure, provide a quantitative measure of the improvements observed. These results underscore the potential of our approach in applications requiring high-resolution and low-speckle 2D imaging.

\subsection{Experimental Capture - Full Camera View}

\begin{figure*}[tb]
\centering
\includegraphics[width=1.0\textwidth]{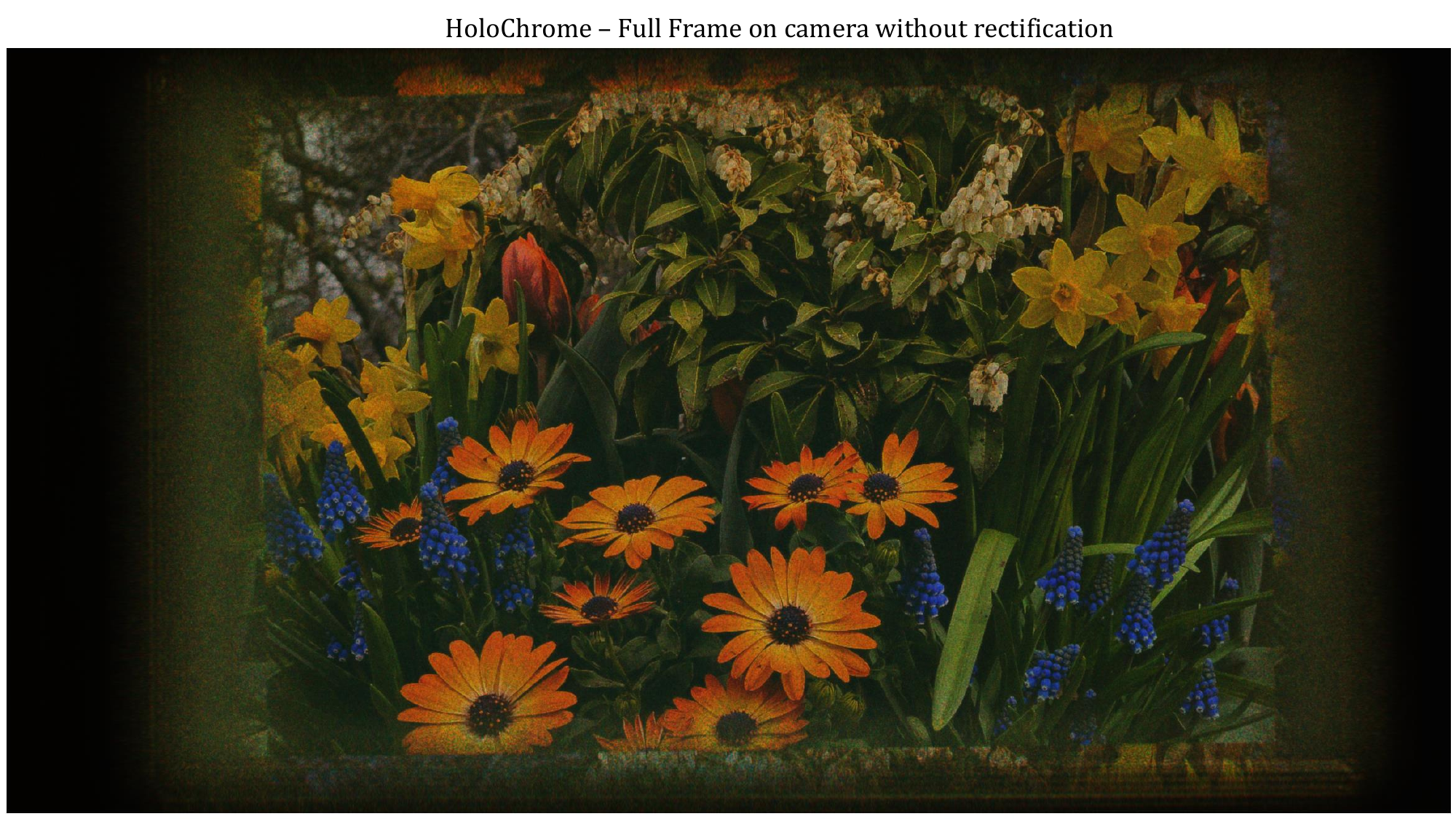}
\caption{
\textbf{HoloChrome 2D Experimental Results without Rectification:}
This figure shows the full-frame HoloChrome 2D experimental result, where 8 individual time-multiplexed frames, each corresponding to a different wavelength, have been combined. The image is captured without any rectification, resulting in the uncorrected appearance shown. The individual time frames are not presented, as they contribute directly to the integrated result.
}
\label{fig:experimental_result_2d_flowers}
\end{figure*}

In the HoloChrome 2D experiment, we capture a full-frame image by time-multiplexing individual monochromatic frames (e.g. 6,8, or 24 frames), each corresponding to a distinct wavelength, ranging from 460 nm to 650 nm. This broad spectrum covers most of the visible light range. Figure \ref{fig:experimental_result_2d_flowers} presents the combined result of this process, showing the unrectified full-frame image, which provides a clear depiction of our experimental setup without applying post-processing alignment or rectification.

To demonstrate the minimal post-processing involved in this experiment, we only convert the monochromatic frames into a single composite color image as outlined in the main paper.
While all earlier results were warped into the SLM frame using a thin-plate spline (TPS) deformation model, the image quality remains largely unaffected without this rectification. 
Notably, wrapping artifacts are visible around the border of the image, as we intentionally cropped the target image in the loss function to avoid boundary and edge effects of the SLM.
With further calibration, it is likely that the full range of the SLM can be utilized.

\subsection{Individual Color Contributions}

\begin{figure*}[tb]
\centering
\includegraphics[width=1.0\textwidth]{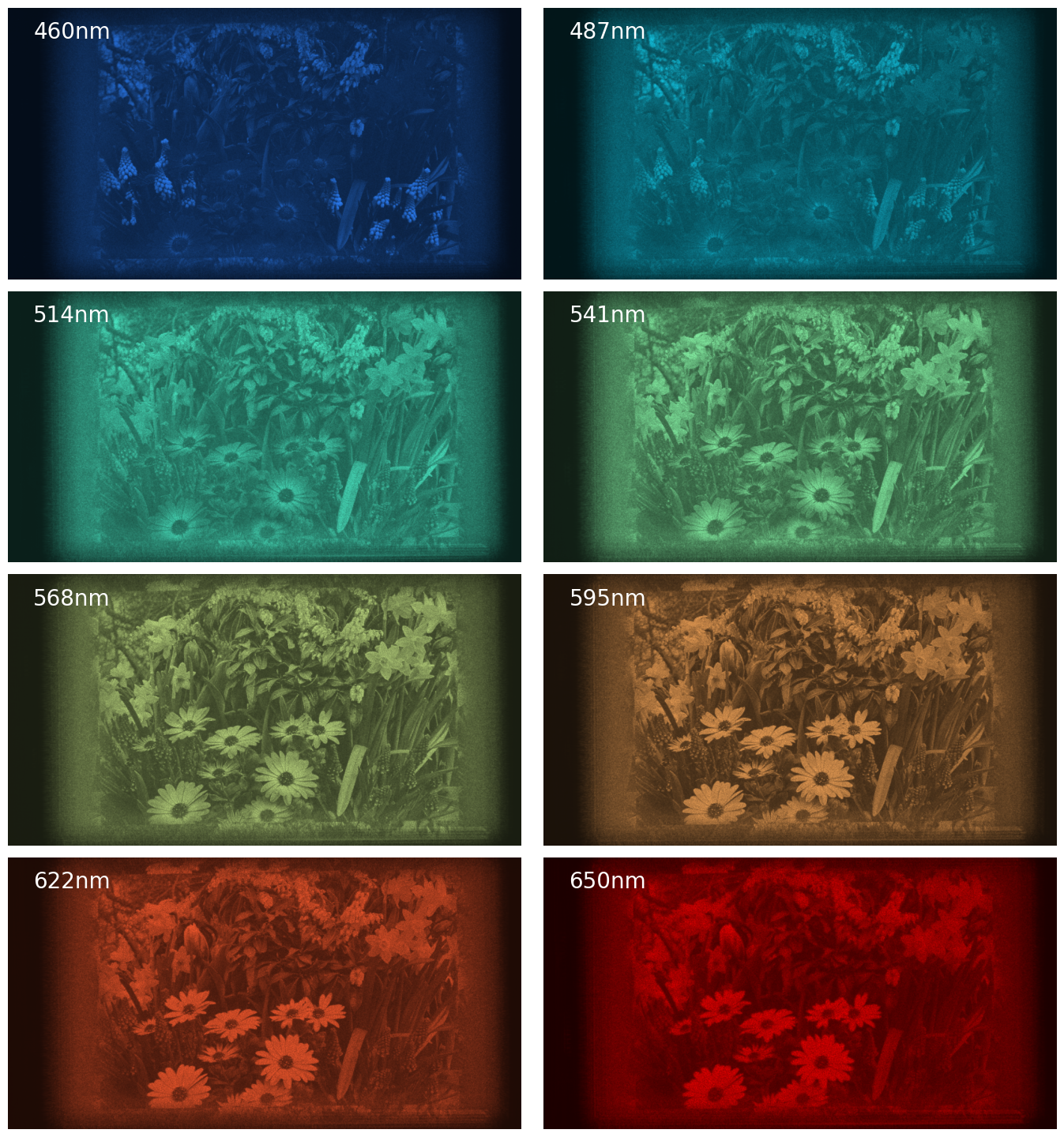}
\caption{
\textbf{Contribution of Monochromatic Responses in HoloChrome 2D Experimental Setup:}
This figure illustrates the contribution of each wavelength to the final composite image in the HoloChrome 2D experimental setup. The full-frame image is captured by time-multiplexing 8 individual monochromatic frames, each corresponding to a specific wavelength (460 nm, 487 nm, 514 nm, 541 nm, 568 nm, 595 nm, 622 nm, and 650 nm). The individual frames are not displayed separately as they are integrated to form the final image.
}    
\label{fig:experimental_result_2d_flowers_indivudal_frames}
\end{figure*}

Figure \ref{fig:experimental_result_2d_flowers_indivudal_frames} provides a detailed look at the contributions of each monochromatic frame to the final composite image. Each frame, corresponding to a specific wavelength (460 nm, 487 nm, 514 nm, 541 nm, 568 nm, 595 nm, 622 nm, and 650 nm), adds a unique component to the overall color result. Note that while this represents a hyperspectral volume, the correct contributions are automatically computed/optimized as we only provide the desired 3D color targets, but not the hyperspectral cube volume in the loss-function.

\section{Model Output vs Experiment}
\label{sec:model_vs_experiment}

In this section, we focus on the comparison between the model output and the experimental captures for both the Time-Sequential and HoloChrome methods. While the overall structure and content of the images align well, there are noticeable discrepancies that warrant further investigation.

\begin{figure*}[tb]
\centering
\includegraphics[width=1.0\textwidth]{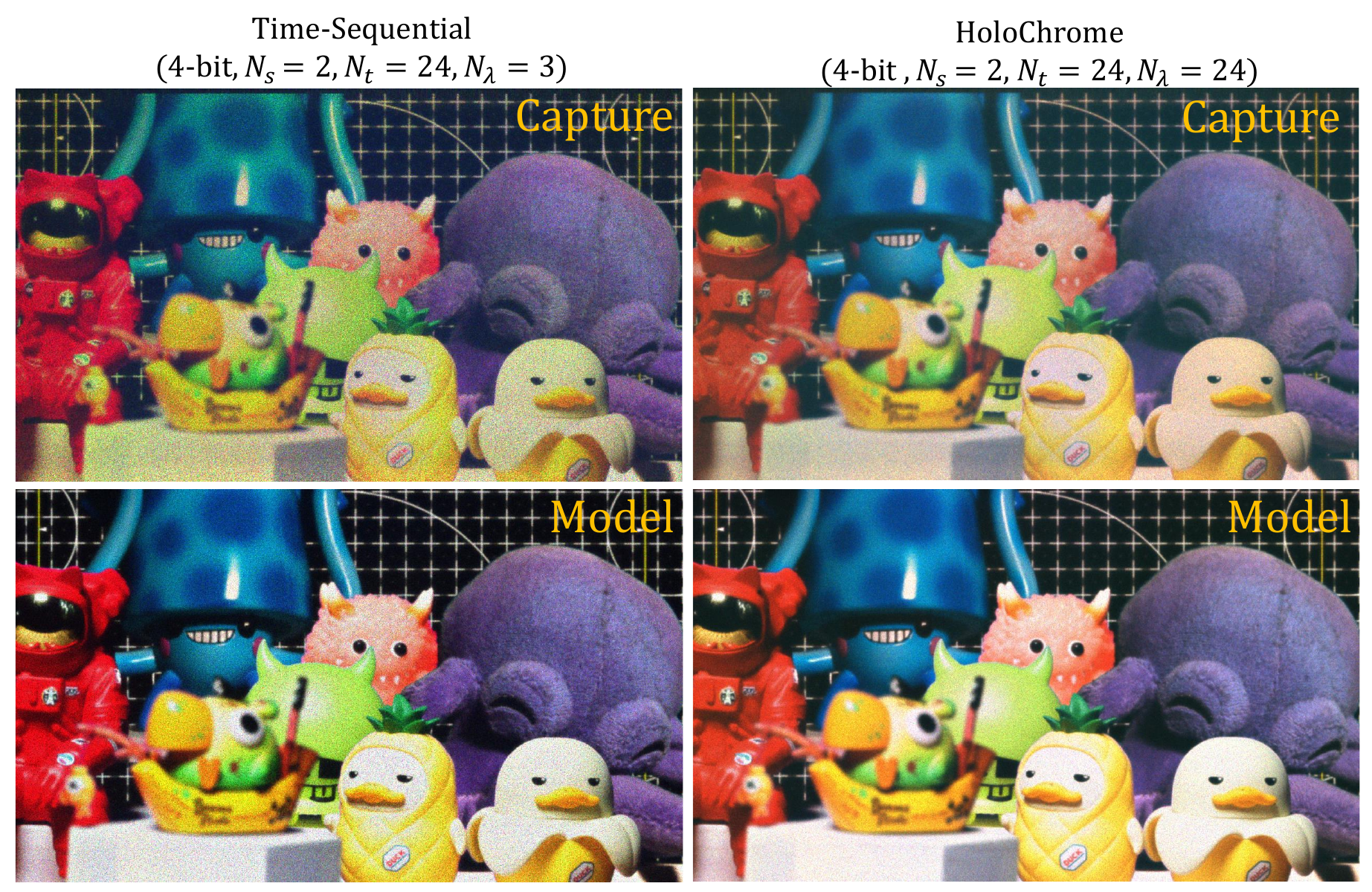}
\caption{
\textbf{Comparison of Model vs. Captured Images for Time-Sequential and HoloChrome Methods:}
This figure compares the experimental captures and the model predictions for both the Time-Sequential and HoloChrome methods. The Time-Sequential method (left column) uses fewer wavelengths (\( N_\lambda = 3 \)), leading to visible discrepancies between the model and captured image, particularly in terms of speckle and contrast preservation. The HoloChrome method (right column) with 24 wavelengths (\( N_\lambda = 24 \)) shows better speckle reduction and higher image fidelity. Despite the overall success of the HoloChrome model in predicting key image features, some inconsistencies between model output and captured images, such as slight variations in black-level reconstruction and contrast, remain.
}    
\label{fig:model_vs_experiment_toys}
\end{figure*}

For the Time-Sequential method, which uses fewer wavelengths (\( N_\lambda = 3 \)), the captured image reveals a higher level of speckle noise and a reduction in contrast when compared to the model output (see Fig.~\ref{fig:model_vs_experiment_toys}). The model accurately predicts the overall intensity distribution but fails to capture the full extent of speckle, particularly in regions with low black levels. This discrepancy may be due to the model's limited ability to simulate real-world optical imperfections and alignment errors during capture.

In the case of the HoloChrome method, which employs a more robust \( N_\lambda = 24 \), the agreement between the model and experimental results is generally better. The model accurately predicts speckle reduction and higher contrast, as seen in both the captured and simulated images. However, some minor differences persist, particularly in black-level reconstruction and fine details at high frequencies. These discrepancies might stem from factors such as unmodeled aberrations, inaccuracies in calibration (as noted earlier in the text), or slight imperfections in experimental setup and laser power control.

Additionally, while our model incorporates a variety of calibration steps (e.g., SLM calibration, alignment between captured images), subtle experimental artifacts can still impact the final image quality. For instance, errors in laser power calibration or optical misalignments could lead to deviations in predicted intensity and contrast. Addressing these factors through improved calibration techniques and refining the optical model could bridge the remaining gap between model predictions and experimental captures.

Overall, the comparison reveals that the model does an excellent job in predicting general trends in image quality—such as speckle reduction and contrast improvement—but also highlights areas where further work is required to fully capture the nuances of the experimental setup.

\subsection{Uniform Eyebox with HoloChrome}

A significant feature of our HoloChrome approach is the uniformity of the eyebox. This is particularly crucial for near-eye display applications to ensure natural accomodation cues.

\begin{figure*}[ht]
    \centering
    \includegraphics{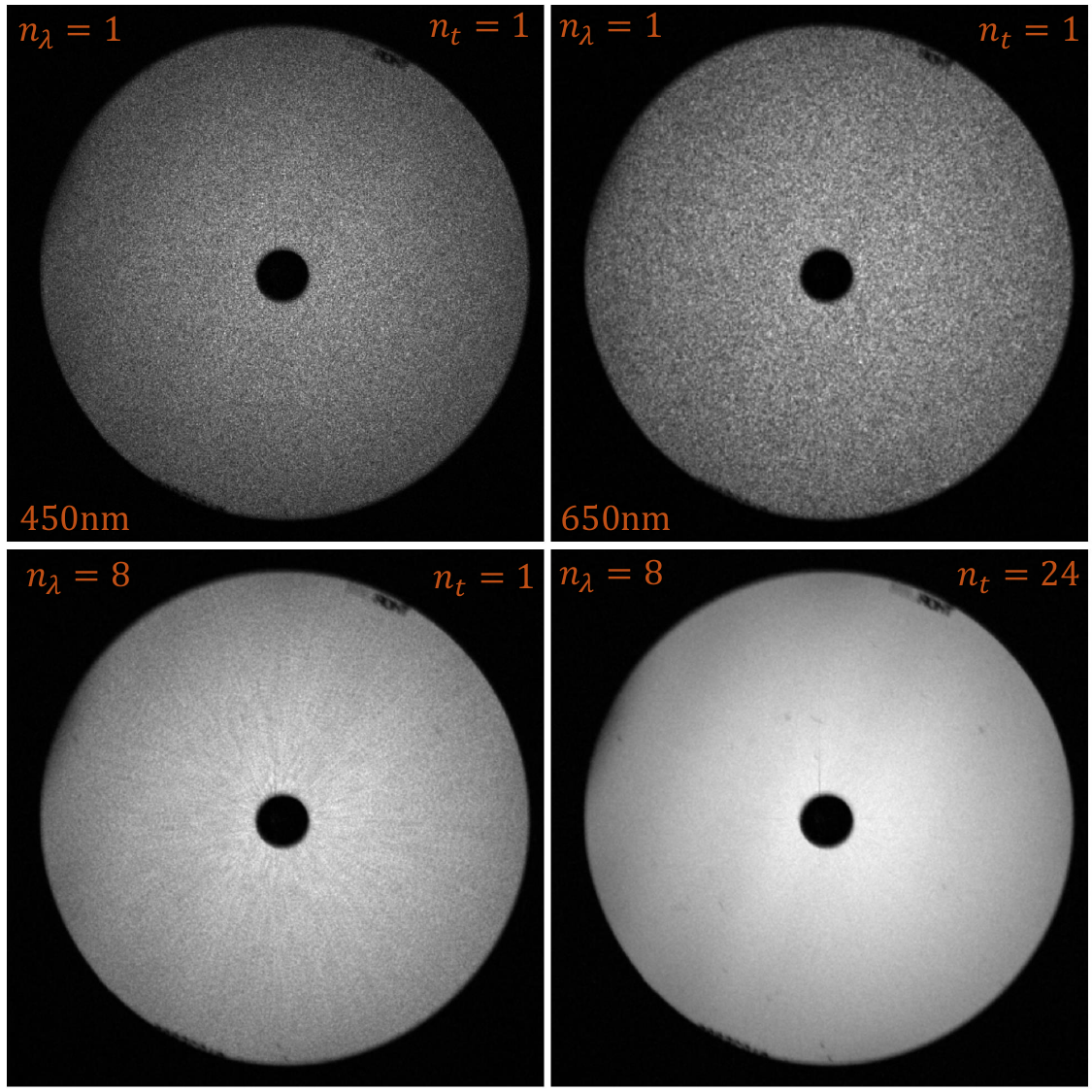}
    \caption{Eyebox intensity for various wavelengths and time frames. The top row shows single-wavelength, single-time-frame results at 450 nm and 650 nm, while the bottom row demonstrates the improvement in uniformity with multi-wavelength ($n_\lambda=8$) and time-multiplexed ($n_t=24$) averaging.
    This is the eyebox for the toys scene used in the main paper.}
    \label{fig:eyebox}
\end{figure*}

In Figure~\ref{fig:eyebox}, the top row presents the eyebox for a single time frame ($n_t=1$) and single wavelength ($n_\lambda=1$) at 450 nm and 650 nm. As can be seen, the eyebox is fully filled, indicating uniform intensity distribution, though some speckle is present. This speckling is expected given the coherence of the system in these single-frame images. Notably, these images are not in the log-domain, so the brightness distribution reflects the actual intensity across the field.

Despite the speckle, even within a single time frame, the averaged result overall wavelengths already shows a trend towards uniformity. The improvement becomes more evident when we average over multiple wavelengths ($n_\lambda=8$) and time-multiplex frame ($n_t=24$), as shown in the bottom right of Figure~\ref{fig:eyebox}. The eyebox becomes exceedingly smooth, demonstrating the progression towards an incoherent display system with enhanced depth cues. The smoothness indicates that the time-averaged system is approaching full accommodation of depth cues, which is a critical advantage for near-eye displays. This temporal and spectral averaging significantly mitigates coherence-related artifacts, making HoloChrome a robust solution for high-quality holographic display systems.

\end{document}